\documentclass[aps,twocolumn,superscriptaddress,nofootinbib]{revtex4}
\usepackage{feyn}
\usepackage{graphicx}
\usepackage{latexsym}
\usepackage{amssymb}
\usepackage{amsmath}
\usepackage{amsfonts}
\usepackage{bm}
\usepackage{multirow}
\usepackage{color}
\usepackage{xcolor}
\usepackage{comment}
\usepackage{bbm}
\usepackage{marvosym}
\usepackage{wasysym}
\usepackage{enumerate}
\usepackage[pdftex,colorlinks=true,linkcolor=blue,citecolor=blue,urlcolor=violet]{hyperref}
\newcommand{\ii}{\mathrm{i}}

\newcommand{\bra}[1]{\left\langle #1\right|}
\newcommand{\ket}[1]{\left|#1 \right\rangle}

\newcommand{\Tr}{\mathop{\mathrm{Tr}}}

\newcommand{\refcite}[1]{Ref.\,\onlinecite{#1}}
\newcommand{\eqnref}[1]{Eq.\,\eqref{#1}}
\newcommand{\figref}[1]{Fig.\,\ref{#1}}

\newcommand{\secref}[1]{Sec.\,\ref{#1}}
\newcommand{\appref}[1]{Appendix~\ref{#1}}

\newcommand{\beq}{\begin{equation}}
\newcommand{\eeq}{\end{equation}}
\newcommand{\beqn}{\begin{eqnarray}}
\newcommand{\eeqn}{\end{eqnarray}}

\DeclareMathAlphabet{\mathbbold}{U}{bbold}{m}{n}

\def\Tr{\text{Tr}}

\newcommand{\Z}{\mathbb{Z}}

\usepackage{tcolorbox}

\begin{document}
\title{Average-exact mixed anomalies and compatible phases}
\author{Yichen Xu}
\author{Chao-Ming Jian}
\affiliation{Department of Physics, Cornell University, Ithaca, New York 14850, USA}

\begin{abstract}
    The quantum anomaly of a global symmetry is known to strongly constrain the allowed low-energy physics in a clean and isolated quantum system. However, the effect of quantum anomalies in disordered systems is much less understood, especially when the global symmetry is only preserved on average by the disorder.   
    In this work, we focus on disordered systems with both average and exact symmetries $A\times K$, where the exact symmetry $K$ is respected in every disorder configuration, and the average $A$ is only preserved on average by the disorder ensemble. When there is a mixed quantum anomaly between the average and exact symmetries, we argue that the mixed state representing the ensemble of disordered ground states cannot be featureless. While disordered mixed states smoothly connected to the anomaly-compatible phases in clean limit are certainly allowed, we also found disordered phases that have no clean-limit counterparts, including the glassy states with strong-to-weak symmetry breaking, and average topological orders for certain anomalies. We construct solvable lattice models to demonstrate each of these possibilities. We also provide a field-theoretic argument as a criterion for whether a given average-exact mixed anomaly admits a compatible average topological order.
\end{abstract}

\maketitle

\tableofcontents
\section{Introduction}
Global symmetry and its quantum anomaly exert significant constraints on the physics of quantum many-body systems. Originally introduced in the context of quantum field theory \cite{adler1969axial,bell1969pcac,hooft1980naturalness}, modern theoretical development has demonstrated the omnipresence of quantum anomalies in various condensed matter systems. For example, on the boundary of a symmetry-protected topological (SPT) state, anomalous symmetry actions are the manifestation of the quantum anomaly induced by the topologically non-trivial bulk \cite{gu2009tensor,chen2012symmetry,chencohomology, else2014classifying,kapustin2014anomalies,witten2016parity,tiwari2018bosonic,kawagoe2021anomalies}. Another class of example is given by the lattice systems subject to the celebrated Lieb-Schultz-Mattis-Oshikawa-Hastings (LSMOH) theorem \cite{lsm,oshikawaluttinger,hastingslsm} and generalizations \cite{chengtranslation,metlitskianomaly,cholsm,jianlsm,chengseiberg,jianglsm,else2020topological}, where mixed quantum anomalies between the crystalline symmetries (translation, crystal rotation, and etc) and the internal symmetries (such as spin rotations) emerge at low-energies. Quantum anomalies, once present, are topologically robust and strongly constrain the allowed quantum phase in the IR. In particular, they forbid the quantum ground states to be ``trivially gapped" regardless of the details of the Hamiltonian. The ground state compatible with the quantum anomaly must be either gapless, symmetry-breaking, or topologically ordered.

The aforementioned situations mostly concern the pure ground states of clean and isolated quantum systems. Meanwhile, physical systems that contain random disorders or interact with an open environment are naturally described by mixed (quantum) states. A mixed state is a probabilistic mixture of an ensemble of pure states. In the case of disordered systems, the relevant mixed state is formed by the ensemble of ground states in different disorder configurations. From the symmetry perspective, in addition to exact symmetries that leave each individual pure state invariant, a mixed state can also be symmetric under average symmetries that map the pure states within the ensemble to one another (A more formal definition of the exact and average symmetries will be given later). Our understanding of the implication of quantum anomalies of exact and/or average symmetries in mixed quantum states is still in an early stage. Refs. \onlinecite{hastingsdisorderlsm, fudelocalization,kimchirandommagnet,ASPT1} investigated the disordered anomalous systems with average spatial symmetries and found the anomaly (or, equivalently, the disordered version of the LSMOH theorem) to impose significant constraints on the low-energy physics still. For example, Ref. \onlinecite{kimchirandommagnet} argued that the ground-state ensemble of disordered magnets with spin 1/2 per statistical unit cell cannot be constructed using the finite-depth quantum circuit. 
For anomalies in open quantum systems \cite{zhou2023reviving,hsin2023anomalies,kawabata2024lieb,lessa2024mixed,wang2024anomaly}, the anomaly-induced constraints are mostly characterized by various entanglement or dynamical aspects of the mixed state.  

In contrast to the existing literature, our focus in this work is to identify possible phases that are compatible with the mixed anomaly between an average symmetry and an exact symmetry, dubbed the \textit{average-exact mixed anomaly} (AEM anomaly), in disordered systems.
In these systems, each realization of the disorders preserves the exact symmetry but breaks the average symmetry explicitly. However, different disorder realizations are related to each other by the action of the average symmetry, making the average symmetry preserved by the entire disorder ensemble. Most crucially, apart from the possible phases that smoothly connect to the clean limit, we discover mixed-state phases that are compatible with the AEM anomaly but do not have clean-limit counterparts. To demonstrate these new possibilities, we construct solvable lattice models for each of these cases. These constructions can be easily generalized to AEM anomalies of other global symmetry groups. We further use field-theoretic arguments to understand what type of AEM anomalies allow these new possibilities.

For clarity, we denote the entire global symmetry group by $G=A\times K$, where $A$ is the average symmetry group and $K$ is the exact symmetry group by our convention. We will subsequently use the term \textit{exact anomaly} when the entire symmetry group associated with the anomaly are exact symmetries.  

Below, we list the AEM anomaly-compatible disordered phases that we have found in this work and define their diagnostics via various forms of correlators of the order parameters in the average density matrix over all the disorder realizations. We note that the list below is not exhaustive, as we do not attempt to fully classify all possible disordered ground states of local Hamiltonians that are compatible with the AEM anomaly.

\begin{itemize}
      \item \textit{Spontaneous symmetry breaking (SSB) states} - 
     Analogous to the clean limit, the AEM anomaly is compatible with a state that spontaneously breaks any part of the symmetry group $G$ in the conventional sense. Such SSB states feature long-ranged correlation functions in the disorder-averaged mixed density matrix $\rho$
     \begin{equation}
        \lim_{|x-y|\to\infty}\Tr[\rho O(x)O^\dagger(y)]\sim O(1)
    \end{equation}
    for the order parameter $O(x)$ that transforms non-trivially under $G$.

    \item \textit{ Glassy states with strong-to-weak spontaneous symmetry breaking (SWSSB)} -
    In \secref{sec:stw}, we show that there exists a different type of spontaneous symmetry-breaking state, known as states with SWSSB, that is also compatible with the AEM anomaly. In such a SWSSB state, the exact symmetry $K$ is spontaneously broken down to an average symmetry, instead of being broken down completely to a trivial group \cite{leeqcpdecoherence,ASPT2,swssb,sala2024spontaneous,zhang2024fluctuation}. In the open quantum systems where SWSSB were introduced, the exact/average symmetries are also referred to as strong/weak symmetries and, hence, the name strong-to-weak SSB.
    This recently proposed SWSSB possesses glassy orders that can be diagnosed by an exponentially decaying correlator of an order parameter $O(x)$ that transforms non-trivially under the exact symmetry $K$,
    \begin{equation}
        \Tr[\rho O(x)O^\dagger(y)]\sim e^{-|x-y|/\xi},
    \end{equation}
    but a non-vanishing long-range R\'enyi-2 correlator
    \begin{equation}
        \lim_{|x-y|\to\infty}\frac{\Tr\left[\rho O(x)O^\dagger(y)\rho O(y)O^\dagger(x)\right]}{\Tr\left(\rho^2\right)}\sim O(1).
    \end{equation}  
    Note that another form for long-range correlator (known as the ``fidelity correlator") has been proposed as an indicator for SWSSB \cite{swssb}. It is our working assumption that the R\'enyi-2 correlator can be used as the diagnosis for SWSSB. We will comment on the fidelity correlator when suitable. 

    \item \textit{Topological and average topological orders} - For {\it exact} anomaly in spatial dimensions $d\geq 2$,  topological orders (TO) with anomalous symmetry actions on the anyons offer a possible option for quantum phases compatible with the anomaly\cite{chongwangbspt,kravec2015all,chensemion,metlitski2015symmetry,wang2017anomaly,qi2019folding,barkeshliset}. For the AEM anomaly, we show that the presence of disorder in TO leads to two options. One option is a disordered TO smoothly connected to the clean limit-TO with the anomalous symmetry action. In some cases, there is a new option, an average topological order (ATO), which can be thought of as a TO but with partially lost quantum coherence caused by random anyonic excitations. We will discuss the conditions for whether anomaly-compatible ATOs are possible for the AEM anomaly of $G=A\times K$, and demonstrate our result by constructing solvable lattice models in \secref{sec:ato}.

\item \textit{Critical disordered states} - Between these disordered phases, there could exist critical disordered phases that is compatible with the AEM anomaly. We will briefly comment on their potential physical features in \secref{sec:disorderfeature}, and leave the study of these options to future works.

\end{itemize}

\section{Review of the exact anomaly}
\subsection{Anomalous symmetry actions on the boundary of SPT states}
\label{sec:anomaly}
In this section, we review how a mixed anomaly of two {\it exact} global symmetries $A$ and $K$ naturally arises at the boundary of SPT states in the clean limit \cite{else2014classifying,kawagoe2021anomalies}. This brief review will set the stage for the discussion of the AEM anomalies in disordered systems in the subsequent sections. We use $d$ to denote the boundary's spatial dimension. The corresponding bulk SPT hence lives in $d+1$ spatial dimensions or $(d+2)$ spacetime dimensions. 

The SPT with exact symmetry $G=A\times K$ that leads to the desired exact mixed anomaly on its boundary can be formulated via the decorated defect construction of SPT states.\cite{xiechenddw} This construction starts with a system where $K$ is preserved, but $A$ is spontaneously broken. This system can contain different configurations of topological defects, such as domain walls, domain-wall junctions, etc, associated with the global symmetry $A$. Each $d_A$ dimensional $A$-defect can be decorated by a non-trivial SPT state in the same dimension protected by the symmetry $K$.  If $K$ is trivial, the ``decoration" is simply a $U(1)$ phase that depends on the $A$-defect configuration. 
Then, one can proliferate the defects of $A$ to restore the full global symmetry $G$. The outcome of the proliferation is a $G$-SPT state that corresponds to a non-trivial element in ${\cal H}^{d-d_A+1}[A,{\cal H}^{d_A+1}[K,U(1)]]$ which, in light of the K\"unneth formula, is a part of the cohomology classification ${\cal H}^{d+2}[A\times K,U(1)]$ of $G$-SPTs.\cite{xiechenddw}.

On the boundary of this $G$-SPT, the bulk $A$-defects, together with the decorated $K$-SPT, can terminate, causing the global symmetry $G$ to act anomalously on the boundary. This is the manifestation of the (exact) mixed anomaly between $A$ and $K$. To understand this anomaly, notice that each $A$-defect that terminates on the boundary hosts a lower-dimensional anomaly of $K$ originating from the boundary of the $K$-SPT state decorated onto the $A$-defect. Consequently, the symmetry action of $K$ is conditioned under $A$-spin configurations, implying that it cannot be strictly local. For a finite group $G$, the explicit form of anomalous symmetry action can be regularized on a lattice based on the element in the cohomology group ${\cal H}^{d+2}[G,U(1)]$, see App. \ref{app:cohomology} which classifies the types of anomalies. The regularized symmetry action is not ``on-site", i.e. not a tensor product of actions on each lattice site, which is a consequence of the anomaly.

As an example useful in later discussions, we consider an anomaly of $G=A\times K$ in $d=2$ where $A=K=\Z_2$. Its corresponding bulk SPT state in 3+1$D$ is constructed by decorating each Ising domain wall of $A=\Z_2$ by a 2+1$D$ SPT state protected by $K=\Z_2$, known as the Levin-Gu state\cite{levingu}. Therefore, in the 2+1$D$ anomaly on the boundary, each domain wall of $A$ hosts a 1+1$D$ $K$-anomaly. The schematic form of the action of $K$ can be written down as follows. We define Ising spins that transform under $A$ and $K$ as $\sigma$ and $\tau$, respectively. Using the decorated defect picture, we can write down the $K$ action localized on a boundary domain wall $DW_\sigma$ of $\sigma^z$, which is inherited directly from the $\Z_2$ symmetry action on the boundary of the Levin-Gu state:
\begin{equation}
    U_K\Big|_{DW_\sigma}=(-1)^{\frac{|DW_\tau\cap DW_\sigma|}{2}}\prod_{i\in DW_\sigma}\tau^x_i,
    \label{ukschematic}
\end{equation}
where $|DW_\tau\cap DW_\tau|$ is the number of intersections with the $K$-domain walls, i.e. the domain wall of the $\tau$ spins, on the boundary $A$-domain wall $DW_\sigma$. Meanwhile, the symmetry action of $A$ is $U_A=\prod_i\sigma^x_i$. The explicit forms of $A$ and $K$ symmetry actions regularized on a square lattice can be found in \secref{sec:z22model} and \appref{app:cohomology}.

The anomaly above corresponds to one of the $\Z_2$ elements in the cohomology group ${\cal H}^4[A\times K =\Z_2^2,U(1)]=\Z_2\times\Z_2$ that classifies the anomaly of $A\times K =\Z_2^2$ in 2+1$D$ (see App. \ref{app:cohomology}). The other $\Z_2$ element in the cohomology classification represents another anomaly where the role of $A$ and $K$ is exchanged, i.e. it arises on the boundary of another 3+1$D$ SPT state where each domain wall of $K$ is decorated with a $\Z_2$-SPT state protected by $A$. For clarity, we denote these two mixed anomalies between $A=K=\Z_2$ as $\Omega_1$ and $\Omega_2$, respectively.

\subsection{Possible physical features of the exact anomaly}
\label{sec:cleanfeature}
We now summarize the physical features of anomaly-compatible phases in the clean limit. The phases we are referring to in this section are all pure states.

It is well-known that the exact anomaly is not compatible with a trivially gapped ground state. Non-trivial quantum phases compatible with the (exact) quantum anomaly include:
\begin{itemize}
    \item SSB phase,
    \item gapless phase,
    \item symmetrically gapped and topologically ordered phase.
\end{itemize}

For a mixed anomaly between $A$ and $K$, it is compatible with a symmetry-breaking state that breaks one of the two subgroups. For example, deep in the SSB phase of $A$, creating an $A$-domain wall is energetically costly. As a result, the symmetry action of $K$ is now essentially anomaly-free without the presence of $A$-domain walls. Hence, the system can still be symmetric under $K$.

There are two alternative scenarios where the global symmetry is not spontaneously broken. One is a gapless state. Possible lattice Hamiltonians of the gapless state can be constructed by symmetrizing the parent Hamiltonian of a trivial product state using the lattice regularization of the anomalous global symmetry actions\cite{dupontpivot,natpivot1,natpivot2}, which usually works for anomalies in 1+1$D$\cite{levingu,zhang2023exactly}. In higher dimensions, numerical methods are usually needed to determine whether a lattice Hamiltonian that is symmetric under the anomalous symmetry action possesses a gapless low-energy spectrum.

For $d\geq 2$, the other anomaly-compatible phase is a topologically ordered state, which will be reviewed in Sec. \ref{sec:SET}.

\section{Disordered mixed state and AEM anomaly}
\subsection{Mixed states and their symmetries}

Physically, a mixed state is a statistical ensemble of pure quantum states $\ket{\psi_\alpha}$ labeled by some index $\alpha$, each with a probability $p_\alpha$ that satisfies $\sum_\alpha p_\alpha=1$. We can describe the mixed state using a density matrix
\begin{equation}
    \rho=\sum_\alpha p_\alpha \ket{\psi_\alpha}\bra{\psi_\alpha},
\end{equation}
so that the average value of an operator $O$ in the mixed state can be conveniently written as $\langle O\rangle\equiv\Tr[\rho O]=\sum_\alpha p_\alpha\bra{\psi_\alpha}O\ket{\psi_\alpha}$.

A mixed state can admit two distinct types of symmetries: exact and average symmetries. Conceptually, a mixed state possesses an exact symmetry $K$ and an average symmetry $A$ if each pure state in the statistical ensemble is symmetric under $K$ (i.e., $U_k\ket{\psi_\alpha}=\ket{\psi_\alpha}$ for each $k\in K$), and any two states related by symmetry $A$ have equal probability (i.e. $p_\alpha=p_\beta$ if $\ket{\psi_\beta}=U_a\ket{\psi_\alpha}$). More generally, a mixed-state density matrix $\rho$ is symmetric under the exact $K$ symmetry and average $A$ symmetry if 
\begin{equation}
    U_k\rho=\rho U_k^\dag=e^{\ii\theta_k}\rho\text{ and }U_a\rho U_a^{\dag}=\rho
    \label{eq:Gsym_Def}
\end{equation}
for all $k\in K$ and $a\in A$. Here, $U_k$ and $U_a$ are the operators that implement the corresponding symmetry actions, and $e^{\ii\theta_k}$ is the corresponding eigenvalue of the exact symmetry $k$.

In this work, we are mainly interested in the case where the global symmetry group is $G=A\times K$, i.e. the average and exact symmetries are independent of each other, and their actions commute. \footnote{More generally the global symmetry group can be a group extension of $A$ by $K$ through the short exact sequence
\begin{equation}
    1\to K\to G\to A\to 1,
\end{equation}
since an average symmetry can be extended to an exact symmetry. For example, there exists a mixed state with global $G=\Z_4=\{1,g,g^2,g^3\}$ symmetry, such that the average symmetry is $g$ and $g^3$ while the exact symmetry is $g^2$. It cannot be the other way round, since if $g^2$ is an average symmetry, $g$ has to also be an average symmetry.}

There are two common settings where such $G$-symmetric mixed states occur. One scenario concerns systems with random quenched disorders. The Hamiltonian of such a disordered system can be written as $H(\{h_i\})$, where $\{h_i\}$ parameterizes the strength of the {\it independent} random local interactions in the vicinity of each lattice site $i$ and follows the probability distribution $p(\{h_i\})$. The mixed state of interest in this scenario is the ensemble of ground states of $H(\{h_i\})$ denoted by $\ket{\psi(h_i)}$. The disordered system is symmetric under $G$ if
\begin{enumerate}[(i)]
    \item the Hamiltonian $H(\{h_i\}$ commutes the $K$ symmetry action for each configuration $\{h_i\}$, namely \beqn U_k H(\{h_i\})U_k^{\dag}=H(\{h_i\})\eeqn for all $k\in K$,
    \item[and]  
    \item the disorder configurations related by the $A$ symmetry action share the same probabilities distribution, namely \beqn U_a H(\{h_i\})U_a^{\dag}=H(\{h^a_i\}),\eeqn such that $p(\{h_i\})=p(\{h^a_i\})$ for all $a\in A$. Here, $\{h^a_i\}$ is the configuration obtained from the symmetry action of $a\in A$ on the configuration $\{h_i\}$. 
\end{enumerate}

Another setting for $G$-symmetric mixed states to occur is an open system. When the $K$ symmetry charge is strictly preserved in the physical system and 
the $A$ symmetry charge can be exchanged between the physical system and the environment, the density matrix of the physical system can also satisfy the condition \eqnref{eq:Gsym_Def}. In the open system context, the exact and average symmetry are also referred to as the strong and weak symmetries. In this work, we will only focus on the mixed state arising from the previous setting of disordered systems.

\subsection{Average SPT state in disordered systems}
\label{sec:ASPT}
Similar to the clean limit, a natural setting that gives rise to the AEM anomaly of $G=A\times K$ is the boundary of the recently proposed average SPT (ASPT) states protected by an average symmetry $A$ and an exact symmetry $K$\cite{ASPT1,ASPT2}. Therefore, we briefly review the ASPT state in disordered systems which helps our subsequent discussion of the AEM anomaly.

A physical scenario of ASPT mixed states is a clean SPT state protected by $G=A\times K$ exposed to random field disorder of the $A$ spins, i.e. the local degrees of freedom that transform non-trivially under $A$. \footnote{An alternative origin for ASPT states is a pure SPT state protected by $G=A\times K$ exposed to measurement or quantum decoherence whose quantum channel possesses strong $K$ and weak $A$ symmetries\cite{ASPT1,zhang2022strange,ASPT2,leesptdecoherence}. } The Hamiltonian for the disordered ASPT state can be written as \beqn H_\text{ASPT}=H_\text{SPT}+H_\text{disorder}(\{h_i\}),\eeqn where $H_\text{SPT}$ denotes the parent Hamiltonian of the SPT state in the clean limit, and $H_\text{disorder}$ contains a set of random ``Zeeman field" $\{h_i\}$ that polarizes the $A$ spins.

The physical picture of the ASPT mixed state is most pronounced in the strong random field disorder limit, where each $A$ spin on lattice site $i$ is polarized along the direction of the random field $h_i$ on the lattice site\footnote{From the Imry-Ma argument\cite{imryma,aizenman1989rounding,aizenman1990rounding} and its subsequent generalization to disorders in quantum systems \cite{greenblattrounding,aizenmanproof,vojtadisorder}, if the spatial dimension of the system is at or below the lower critical dimension $d_c$ ($d_c=2$ for discrete symmetry and $d_c=4$ for continuous symmetry), any spin orders associated with the SSB of $A$ in the clean limit will be unfavorable with the presence of random field disorders that couples to the order parameter of $A$.}. Each ground state of the disordered Hamiltonian is simply given by a configuration of (uncorrelated) $A$ spins pinned by the random field $\{h_i\}$ where each defect in the $A$ spin configuration, including the domain walls, the domain wall junctions, and etc, hosts a lower-dimensional SPT state protected by the exact symmetry $K$. 

With such a picture, we see that the exact symmetry group $K$ has to be non-trivial in order for the ASPT to be topologically non-trivial. An ``ASPT mixed state" with only an average symmetry (i.e. $G=A$) is topologically trivial, since the decoration on the $A$-defects are simply $U(1)$ phases in the clean limit, which lose meaning in a mixed state without coherent superposition of $A$-defects. Mathematically, this means that when calculating the cohomology classification of $G$-ASPT using the K\"unneth formula ${\cal H}^{d+2}[A\times K,U(1)]=\bigoplus_{D=0}^{d+2}{\cal H}^{d+2-D}[A, {\cal H}^D[K,U(1)]]$, the element with $D=0$ no longer enters the classification.\footnote{Ref. \onlinecite{ASPT2} pointed out that the element with $D=1$ also does not enter the classification of disordered ASPT states, since the $K$ charges that are decorated onto the point defects of $A$ can simply localize under disorders. However, as we will see in \secref{sec:stw}, such a bulk is no longer an SPT state due to the fact that $K$ has been spontaneously broken down to an average symmetry when the $K$ charges are localized. Therefore, it is still meaningful to discuss an AEM anomaly that corresponds to the element with $D=1$. }

\subsection{Physical pictures of the AEM anomaly: fluctuation versus percolation}
\label{sec:disorderfeature}

\begin{figure}
    \centering
    \includegraphics[width=7cm]{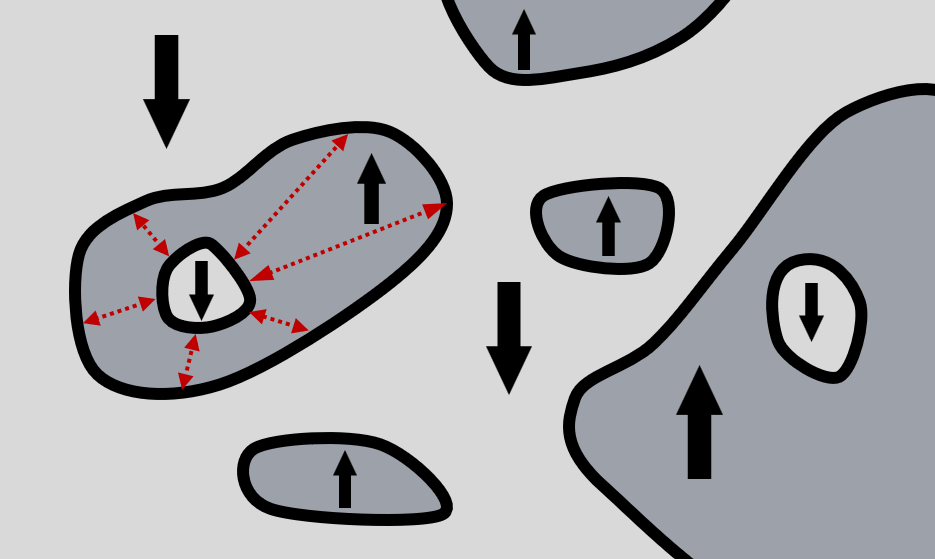}
    \caption{A typical sample in the strong random field disorder limit, where the black arrows indicate the $A$ spins that are quenched and polarized by the strong random fields. On each defect of $A$ marked by the black lines, the exact symmetry $K$ acts anomalously. The red arrows indicates possible interaction that can trivialize two conjugating $K$ anomalies, and the paired $K$ anomaly carries a certain spin of $A$. }
    \label{fig:sample}
\end{figure}

Now that we have established the physical picture of the ASPT state, we can discuss the AEM anomaly between the average symmetry $A$ and the exact symmetry $K$. A way to understand this anomaly is to consider the boundary a strongly disordered ASPT. Similar to the discussions in \secref{sec:anomaly}, the mixed anomaly is manifested in the way that, on each lower-dimensional $A$-defect pinned by strong random-field disorders, the exact symmetry $k\in K$ acts in an anomalous way (see \figref{fig:sample}). In a disordered setting, the $A$-defect configurations are not coherently superposed, but simply mixed together incoherently in the density matrix. As we will see below, this is the main reason that gives rise to anomaly-compatible disordered phases beyond the clean-limit possibilities.

Using such a decorated defect picture, we now discuss possible features of the mixed state that are compatible with the AEM anomaly of $ G=A \times K$. 

We first consider the case where the exact symmetry $K$ is trivial. Recall from Sec. \ref{sec:ASPT} that any ASPT mixed state protected solely by an average symmetry $G=A$ is topologically trivial, we would expect that the anomaly of purely average symmetry cannot prevent the system from becoming completely trivial without any physical feature. Indeed, in the strong disorder limit, the system can be simply a mixture of different quenched $A$-defect configurations, whose density matrix is proportional to identity (i.e. an infinite-temperature state): $\rho \propto \mathbbm{1}$. Such a disordered state will always be symmetric under any average symmetry.  

Moving on to the AEM anomaly of $G=A\times K$ with a non-trivial $K$. Due to the presence of lower-dimensional $K$-anomalies decorated onto the $A$-defects, each of them exhibits some non-trivial physical features. What feature occurs depends on the details of interactions in the system. If the disorder induces a strong localization effect, a possibility is that  $K$ is spontaneously broken on each $A$-defect in order to be compatible with the anomaly of the exact symmetry $K$. The feature of this scenario can be potentially detected using the correlators of the order parameters of $K$. In fact, the behavior of the correlator depends on whether the $A$-defects percolate or not. To see this, we consider the ground state $\ket{\psi_\alpha}$ of some disorder realization $\alpha$ where each $A$-defect is decorated with an SSB state of $K$. Since the $K$ is spontaneously broken on each $A$-defect, we have
\begin{equation}
    O(x)\ket{\psi_\alpha}\sim  O(y)\ket{\psi_\alpha}
\end{equation}
for some local order parameter $O$ of $K$ only if $x$ and $y$ are located on the same $A$-defect. If the defects of $A$ percolate, such a possibility is almost 1 across the entire ensemble\cite{stauffer2018introduction}. Therefore, the entire disordered state breaks $K$ spontaneously, which can be diagnosed by the correlator of the order parameter $O(x)$ averaged over all possible disorder realizations:
\begin{equation}
    \langle O(x)O^\dagger(y)\rangle\equiv\lim_{|x-y|\to\infty}\Tr[\rho O(x)O^\dagger(y)]\sim O(1).
    \label{trrhoOO}
\end{equation}
If the $A$-defects do not percolate, namely, if the SSB of the exact $K$ symmetry remains independent on each $A$-defect, such a correlator will instead be short-ranged. We will show in \secref{sec:stw} that such a situation indicates a breakdown of the exact symmetry to an average symmetry, i.e. an SSWSB of $K$, in contrast to the usual SSB where $K$ is broken down to a trivial group.

On the other hand, if there exists strong enough interaction between $K$-spins, pairs of $A$-defects with conjugating $K$-anomalies can be ``gapped out" while preserving $K$ (see \figref{fig:sample}). However, forming such a pairing of all the $A$ defects will favor a specific direction of $A$-spins. Therefore, when the $A$-defects together with their $K$-anomalies are paired up under interaction, we argue that the average symmetry $A$ will be spontaneously broken.

Between these two limits, there could, in principle, exist a critical disordered state that is compatible with the AEM anomaly.  For example, consider the case where each 1+1$D$ or higher dimensional $A$-defect hosts the ground state of some gapless Hamiltonian that is compatible with the $K$ anomaly, and the $A$-defects percolate. Therefore, the gap of such a disordered system should vanish in the thermodynamic limit for a typical disorder realization, i.e. the system is gapless on average. The physical picture here is conceptually similar to the quantum Hall plateau transition at half-integer filling \cite{chalkercoddington,leeplateau,wangplateau} and the delocalized state at the Anderson localization transition for Dirac fermions \cite{morimoto2015anderson,fudelocalization}. We also note that the correlator within a typical disorder realization is not necessarily power-law in this case but the disordered average correlators are expected to be power-law. One example is the random singlet state \cite{fisherrandom,liu2018random,liu2020quantum} that is compatible with the average LSM anomaly with an exact spin-rotation symmetry $K=SO(3)$ and an average translation symmetry $A$, where the spin-spin correlator decay as $\langle \mathbf{S}(x)\cdot\mathbf{S}(y)\rangle\sim |x-y|^{-2}$ when averaging over all disorder realizations, but $\bra{\psi_\alpha}\mathbf{S}(x)\cdot\mathbf{S}(y)\ket{\psi_\alpha}$ for a typical disorder realization $\alpha$ can be short-ranged. In fact, Ref. \onlinecite{kimchirandommagnet} proved in 1+1$D$ that, in a disordered mixed state that is compatible with the average LSM anomaly, the average spin-spin correlator cannot decay faster than the one for the random singlet phase without spontaneously breaking the average translation symmetry.

We note that there also exist interactions that can lead to a symmetrically gapped state with topological order or average topological order that is compatible with the AEM anomaly, which we discuss in detail in \secref{sec:SET}.

\section{SWSSB states compatible with the AEM anomaly}
\label{sec:stw}

We now consider the case with SWSSB where the $A$-defects do not percolate in more detail. As alluded in \secref{sec:disorderfeature}, even if each $A$-defect hosts an SSB state of $K$ in this case, the two-point correlators of the order parameters of $K$ in Eq. \eqref{trrhoOO} will decay exponentially with relative distance, since these two points have almost-zero probability of being on the same $A$-defect, i.e.
\begin{equation}
    O(x)\rho\not\sim O(y)\rho.
    \label{notsim}
\end{equation}
The physical picture here is that the $K$ symmetry breaking pattern fluctuates independently at each $A$-defect, i.e., the system is a glassy state of $K$-charges. 

To probe the symmetry-breaking order on each $A$-defect, notice that the following similarity holds between the two density matrices with order parameter of $K$ acting at two different spatial locations:
\begin{equation}
    O(x)\rho O^\dagger(x) \sim O(y)\rho O^\dagger(y),
    \label{OrhoO}
\end{equation}
regardless of the positions $x$ and $y$. This is because the $K$ charges are now independently condensed on both the $A$ defects on which $x$ and $y$ are located\footnote{$O(x)\rho O^\dagger(x)$ is not similar to $\rho$, since the entire state ensemble in $\rho$ share a definitive symmetry charge of $K$ different that of $O(x)\rho O^\dagger(x)$.}. As proposed by Refs. \onlinecite{leeqcpdecoherence, ASPT2, swssb}, the relations in Eqs. \eqref{notsim} and \eqref{OrhoO} indicate that the exact symmetry $K$ has been spontaneously broken down to an average symmetry, rather than being broken down completely to a trivial group. Such a pattern of symmetry breaking is called strong-to-weak spontaneous symmetry breaking (SWSSB). To quantify the similarity suggested by \eqref{OrhoO}, a convenient choice to calculate is the normalized overlap between these two density matrices:
\begin{equation}
\langle O(x)O^\dagger(y)\rangle_2\equiv\frac{\Tr[O(x)\rho O^\dagger(x)O(y)\rho O^\dagger(y)]}{\Tr[\rho^2]},
\label{renyi2corr}
\end{equation}
which is the R\'enyi-2 version of the two-point charge correlator of $K$. In short, the SWSSB phase is characterized by a finite value of $\lim_{|x-y|\to\infty}\langle O(x)O^\dagger(y)\rangle_2$ and short-range correlated $\langle O(x)O^\dagger(y)\rangle$\footnote{As pointed out by Ref. \onlinecite{swssb}, a characterization of the similarity in \eqnref{OrhoO} that is more robust against perturbations via strongly symmetric local quantum channels is through the fidelity between these two density matrices: \beqn\langle O(x)O^\dagger(y)\rangle_2^F\equiv F[ O(x)\rho O^\dagger(x),O(y)\rho O^\dagger(y)],\eeqn where $F(\rho,\sigma)=\left[\Tr(\sqrt{\sqrt{\rho}\sigma\sqrt{\rho}})\right]^2$ is the fidelity between two density matrices $\rho$ and $\sigma$. The two correlators are generally inequivalent but behave qualitatively similarly in diagnosing SWSSB order deep inside this phase.}.

We note that the definition in \eqnref{renyi2corr} is related to the Edwards-Anderson order parameter of spin glass \cite{parisiorder}. To see this, we plug in the density matrix with the diagonal basis $\rho=\sum_\alpha p_\alpha\ket{\psi_\alpha}\bra{\psi_\alpha}$ and assume that any two disorder realizations $\alpha$ and $\beta$ have zero overlap even with a local charge excitation, i.e. $\bra{\psi_\alpha}O(x)O^\dagger(y)\ket{\psi_\beta} \approx 0$. We then have
\begin{equation}
    \langle O(x)O^\dagger(y)\rangle_2\approx \frac{\sum_\alpha p_\alpha^2 |\bra{\psi_\alpha}O(x)O^\dagger(y)\ket{\psi_\alpha}|^2}{\sum_\alpha p_\alpha^2},
\end{equation}
which matches the Edwards-Anderson order parameter if $O$ is taken to be the Ising spin operator and the probability for each disorder realization $\alpha$ is proportional to $p_\alpha^2$.\footnote{The fidelity correlator of SWSSB, $\langle O(x)O^\dagger(y)\rangle_2^F$, will be simplified to the Edwards-Anderson order parameter where each disorder realization $\alpha$ has probability $p_\alpha$, instead of $p_\alpha^2$.}

\subsection{A solvable model with SWSSB }
\label{sec:stwmodel}
\begin{figure}
    \centering
    \includegraphics[width=6cm]{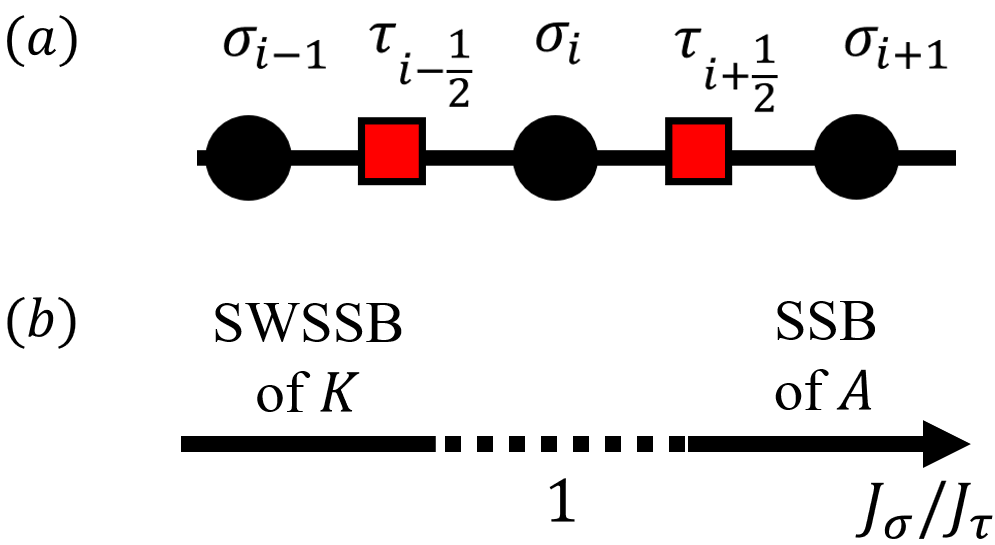}
    \caption{(a) The 1$D$ lattice realization of the mixed anomaly between $A=Z=\Z_2$.(b) Schematic phase diagram of the random Hamiltonian in \eqnref{HJh}.}
    \label{fig:1dlattice}
\end{figure}

To demonstrate a 1+1D SWSSB state that is compatible with the AEM anomaly, we construct a physical disordered model to realize an SWSSB of the exact symmetry $K$ that is compatible with the 1+1$D$ AEM anomaly of $G=A\times K$ between $A=K=\Z_2$. The anomalous symmetry we focus on can be defined on a 1$D$ lattice as the following non-on-site actions (see App. \ref{app:cohomology} for the derivation of the anomalous symmetry actions on the lattice starting from the group cohomology that characterizes the anomaly).
\begin{equation}
    U_A=\prod_{i=1}^N\tau^x_{i+\frac{1}{2}},\ U_K=\prod_{i=1}^N C\left(\sigma^z_i,\tau^z_{i-\frac{1}{2}}\tau^z_{i+\frac{1}{2}}\right)\sigma^x_i,
    \label{UAUK}
\end{equation}
where the $\sigma$ spins are located on the lattice sites and $\tau$ spins are on the bonds (see \figref{fig:1dlattice}), and $N$ is the number of lattice sites. $C(A,B)=e^{\frac{\ii\pi}{4}(1-A)(1-B)}$ is the controlled unitary between two Pauli operators $A$ and $B$. We also assume a periodic boundary condition $\sigma^z_{N+1}=\sigma^z_1$ and $\tau^z_{N+\frac{1}{2}}=\tau^z_{\frac{1}{2}}$. One can verify that $U_K^2=1$. From the expression of $U_K$, the anomaly is manifested in the way that each $\tau^z$ domain wall hosts a half charge of $K$: $U_K\Big|_{\tau^z_{i-\frac{1}{2}}\tau^z_{i+\frac{1}{2}}=-1}=\sigma^z_i\sigma^x_i=\ii\sigma^y_i$, so that locally on each domain wall $U_K$ squares to $-1$. 

In the clean limit (where both $A$ and $K$ are exact), various anomaly-compatible states can be constructed using the following Hamiltonian\cite{xiechenddw}
\begin{equation}
\begin{aligned}
    &H_{\Z_2^2}=\\
    &-\left[\sum_i(J_\sigma\sigma^x_i+J_\tau\tau^x_{i+\frac{1}{2}})+U_K\sum_i(J_\sigma\sigma^x_i+J_\tau\tau^x_{i+\frac{1}{2}})U_K^{-1}\right]\\
    &=-\sum_i\left[J_\tau\tau^x_{i+\frac{1}{2}}\left(1+\sigma^z_i\sigma^z_{i+1}\right)+J_\sigma\sigma^x_i\left(1+\tau^z_{i-\frac{1}{2}}\tau^z_{i+\frac{1}{2}}\right)\right].
\label{Hz22gapless}
\end{aligned}
\end{equation}

 The ground state of the Hamiltonian can be tuned by $J_\sigma/J_\tau$. When $J_\sigma/J_\tau\gg1$ (or $\ll1$), the Hamiltonian favors a ferromagnetic ground state of $\tau^z$ (or $\sigma^z$) while $\sigma^x_i=1$ (or 
 $\tau^x_i=1$) on each site, which spontaneously breaks $A$ (or $K$). These two SSB states are separated by a critical point at $J_\sigma=J_\tau$, at which the Hamiltonian is gapless and is exactly solvable by mapping to a fermionic chain.
 
To construct a disordered state with SWSSB of $K$, we need to design a random-bond type of disorder that enforces a ferromagnetic state of $\sigma^z$ in a ``random basis", while also explicitly breaking the symmetry $A$ down to an average symmetry. The random-basis ferromagnetic can be achieved by defining the following unitary operator that randomly rotate the basis for $\sigma^z$
\begin{equation}
    W(\{h_i,l_i\})=\prod_i V_i^{\frac{1-h_i}{2}}\left(\tau^z_{i+\frac{1}{2}}\right)^{\frac{1-l_i}{2}},
    \label{Whl}
\end{equation}
where $V_i=\sigma^x_iC\left(\sigma^z_i,\tau^z_{i-\frac{1}{2}}\tau^z_{i+\frac{1}{2}}\right)$ and $h_i,l_i=\pm 1$ with equal probabilities denotes random flips required for a SWSSB state. The unitary $W$ is symmetric under $K$ and $A$ since
\begin{align}
    & U_A W(\{h_i,l_i\})U_A^{\dag}=\left(\prod_il_i\right)W(\{h_i,l_i\}), \nonumber \\
     & U_K W(\{h_i,l_i\})U_K^{\dag}=W(\{h_i,l_i\}).
\end{align}
Note that the random basis rotation $W(\{h_i,l_i\})$ randomly adds to the system a symmetry charge $l_i =\pm 1$ of $A$  with equal probabilities in every unit cell. Using this random basis rotation, we construct a disordered Hamiltonian:
\begin{equation}
    H(\{h_i,l_i\})\equiv W(\{h_i,l_i\})H^Y_{\Z_2^2}W^{-1}(\{h_i,l_i\}),
\label{HJh}
\end{equation}
where 
\begin{equation}
\begin{aligned}
    H^Y_{\Z_2^2}=&-\sum_i\left[J_\tau\tau^y_{i+\frac{1}{2}}\left(1+\sigma^z_i\sigma^z_{i+1}\right)\right.\\
    &\quad\quad\quad\quad\quad \left.+J_\sigma\sigma^x_i\left(1+\tau^z_{i-\frac{1}{2}}\tau^z_{i+\frac{1}{2}}\right)\right]
\end{aligned}
\end{equation}
is obtained from $H_{\Z_2^2}$ by modifying the $J_\tau$ interaction so that it explicitly breaks $A$. The ensemble of random Hamiltonians still preserves $A$ on average:
\begin{equation}
    U_A H(\{h_i,l_i\})U_A^{-1}=H(\{h_i,-l_i\}).
\end{equation}
Since $V_i$ commutes with other $V_j$ and $\tau^z_{j+\frac{1}{2}}$, each of the randomly transformed Hamiltonian $H(\{h_i,l_i\})$ still consists of local terms (see \appref{app:derivation} for its explicit form). Physically, the randomly rotated $J_\tau$ term mimics a random bond Ising-type interaction between two nearby spins that is symmetric under $K$ but explicitly breaks $A$.

We now solve for the mixed-state density matrix for the ensemble of ground states of $H(\{h_i,l_i\})$. First, the ferromagnetic ground state of $H^Y_{\Z_2^2}=H(\{h_i=1,l_i=1\})$ in the limit $J_\sigma/J_\tau\ll 1$ reads \beqn\ket{\psi_0}=\frac{1}{\sqrt{2}}\left(\ket{\uparrow\uparrow\dots}_\sigma+\ket{\downarrow\downarrow\dots}_\sigma\right)\otimes\ket{+_y+_y\dots}_\tau,\label{psi0}\eeqn where the $\sigma$ spins are in the superposition of two ordered configurations so that $\ket{\psi_0}$ is overall charge-neutral under $K$. Energetically, we can show that the state $\ket{\psi_0}$ with the combination $\left(\ket{\uparrow\uparrow\dots}_\sigma+\ket{\downarrow\downarrow\dots}_\sigma\right)$ has a lower energy than its counterpart with $\left(\ket{\uparrow\uparrow\dots}_\sigma-\ket{\downarrow\downarrow\dots}_\sigma\right)$ in a finite-size system. This can be seen via a degenerate perturbation theory in the subspace of the two degenerate ground states $\{\ket{\uparrow\uparrow\dots}_\sigma,\ket{\downarrow\downarrow\dots}_\sigma\}$, by treating the $J_\sigma$ term in the Hamiltonian $H^Y_{\Z_2^2}$ as a perturbation. Later, we will use $\ket{\psi_0}$ to construct an ensemble of disordered ground states of $H(\{h_i,l_i\})$ within the {\it charge-neutral sector} of $K$. Conceptually, fixing the charge sector is analogous to a standard finite-temperature canonical thermal ensemble with a fixed charge.

Using $\ket{\psi_0}$, the ensemble of disordered ground states of $H(\{h_i,l_i\})$ can then be described by the following density matrix
\begin{equation}
\begin{aligned}
    \rho_{\Z_2^2}&=\frac{1}{2^{2N}}\sum_{h_i,l_i}W(\{h_i,l_i\})\ket{\psi_0}\bra{\psi_0}W^{-1}(\{h_i,l_i\})\\
    &=
    \frac{\mathbbm{1}+U_K}{2^{2N}}.
\end{aligned}
\label{inftemp}
\end{equation}
The detailed derivation can be found in \appref{app:derivation}. The ensemble of disordered ground state can be regarded as an infinite-temperature canonical ensemble of $K$ charges: every state in the ensemble is overall charge-neutral under $K$ since both $\ket{\psi_0}$ and $W(\{h_i,l_i\})$ are $K$-symmetric.
It is straightforward to check that $\rho_{\Z_2^2}$ has an exact $K$ symmetry and an average $A$ symmetry. The SWSSB of the exact symmetry $K$ in $\rho_{\Z_2^2}$ can be detected by the correlators
\begin{equation}
    \frac{\Tr[\rho_{\Z_2^2}\sigma^z_i\sigma^z_j\rho_{\Z_2^2}\sigma^z_i\sigma^z_j]}{\Tr[\rho_{\Z_2^2}^2]}=1 \text{ and }\Tr[\rho_{\Z_2^2}\sigma^z_i\sigma^z_j]=0
\end{equation}
for any $i\neq j$.\footnote{The fidelity correlator $\langle \sigma^z_i\sigma^z_j\rangle_2^F$ for $\rho_{\Z_2^2}$ is also 1: $\langle \sigma^z_i\sigma^z_j\rangle_2^F=F(\sigma^z_i\rho_{\Z_2^2}\sigma^z_i,\sigma^z_j\rho_{\Z_2^2}\sigma^z_j)=F\left(\frac{\mathbbm{1}-U_K}{2^{2N}},\frac{\mathbbm{1}-U_K}{2^{2N}}\right)=1$} 

Interestingly, by increasing $J_\sigma/J_\tau$ , the random Hamiltonian in \eqnref{HJh} interpolates between SWSSB state of $K$ and a SSB state of $A$, as shown in the phase diagram in Fig. \ref{fig:1dlattice}. To see this, we start from the clean limit ferromagnetic state $\ket{\psi_1}=\frac{1}{\sqrt{2}}\ket{++\dots}_\sigma\otimes\left(\ket{\uparrow\uparrow\dots}_\tau+\ket{\downarrow\downarrow\dots}_\tau\right)$ in the limit $J_\sigma/J_\tau\gg 1$ (we pick the superposition of $\tau$ spins for the same reason as in \eqnref{psi0}). In the presence of disorder $\{h_i, l_i\}$, the ensemble of ground states is still an SSB state of $A$:
\begin{equation}
	\begin{aligned}
		\rho_{\Z_2^2}'=&\frac{1}{2^{2N}}\sum_{h_i,l_i}W(\{h_i,l_i\})\ket{\psi_1}\bra{\psi_1}W^{-1}(\{h_i,l_i\})\\
		=&\ket{++\dots}\bra{++\dots}_\sigma\\
		&\otimes\frac{1}{2}\left(\ket{\uparrow\uparrow\dots}\bra{\uparrow\uparrow\dots}_\tau+\ket{\downarrow\downarrow\dots}\bra{\downarrow\downarrow\dots}_\tau\right).
	\end{aligned}
\end{equation}
We leave the nature of the phase boundary between the $A$-SSB state and $K$-SWSSB state to future studies.

\section{Disordered topological states compatible with the AEM anomaly}
\label{sec:ato}
In this section, we consider the possible topologically ordered (TO) state and average topologically ordered (ATO) state that are compatible with the AEM anomaly of $G=A\times K$ in spatial dimension $d=2$. 
\subsection{Topological order and average topological order in disordered systems}
\label{sec:TOmixed}
Before discussing the compatibility with the AEM anomaly, it is necessary to clarify the notion of TO (and ATO) in disordered systems. Similar to the clean limit, we define TO in a disordered 2+1$D$ system as an SSB state of a set of (emergent) exact discrete 1-form symmetries, where each 1-form symmetry is associated with an Abelian anyon in the TO. The TO can be diagnosed by the expectation value of the 1$D$ loop operator $O_\gamma$ that is charged under the 1-form symmetry, which should decay no faster than perimeter law when averaged over all disorder realizations: $\langle O_\gamma \rangle \equiv\Tr(\rho O_\gamma)\sim e^{-\alpha |\gamma|}$\cite{hastings2005quasiadiabatic}, where $\gamma$ represents the loop and $|\gamma|$ is its length. Such a disordered state can be smoothly connected to the clean-limit topological order. On the contrary, in the unbroken phase of the 1-form symmetry, such an expectation value decays with the area $S_\gamma$ that $\gamma$ encircles: $\langle O_\gamma\rangle\sim e^{-\beta S_\gamma}$ with some constant $\beta$. These two types of behaviors of $\langle O_\gamma\rangle$ are commonly known as ``perimeter law" and ``area law", respectively.

Generalizing the interpretation of TO as an SSB state of 1-form symmetry, we can define the ATO state in disordered systems as an SWSSB state of 1-form symmetry\cite{sohal2024noisy,ellisonaverageTO,1formswssb}. Similar to the diagnosis of SWSSB states of a 0-form symmetry in Sec. \ref{sec:stw}, the ATO exhibits an area law in the ensemble average of the loop operator. However, the exact 1-form symmetry is not completely broken down to a trivial group, which is signaled by the
perimeter law of the R\'enyi-2 loop expectation value\footnote{A fidelity-based loop expectation value can be similarly defined as $\langle O_\gamma\rangle_2^F\equiv F(\rho, O_\gamma\rho O_\gamma^\dagger)$}:
\begin{equation}
    \langle O_\gamma\rangle_2\equiv\frac{\Tr[\rho O_\gamma \rho O_\gamma^\dagger]}{\Tr[\rho^2]}\sim e^{-\alpha|\gamma|}.
\end{equation}
In contrast, a topologically trivial state should exhibit an area law for both $\langle O_\gamma\rangle$ and $\langle O_\gamma\rangle_2$. 
We caution that our diagnosis of ATO via $\langle O_\gamma\rangle$ and $\langle O_\gamma\rangle_2$ should be treated as a working assumption for this paper. The optimal diagnosis of ATOs in general disordered and open systems will be left for future studies.\footnote{There is an important caveat in generalizing such a definition to topologically ordered mixed state in open quantum systems\cite{fanbao,baofan,su2023higher,su2024tapestry,li2024replica,sohal2024noisy,ellisonaverageTO}. In these cases the loop operator expectation value $\langle O_\gamma\rangle$ will always decay in perimeter law across the error threshold of the TO, beyond which the error-corrupted TO cannot be connected to the clean TO via local quantum channels, and should hence be regarded as an ATO\cite{sangmixedstate}. SSB of 1-form symmetry is tricky to define in these cases using loop operator as an order parameter, see \cite{1formswssb}.}

A concrete example of ATO is given by the disordered toric-code state (or $\Z_2$ gauge theory) on a 2$D$ square lattice \cite{tsomokos2011interplay}, whose Hamiltonian reads
\begin{equation}
    H_\text{ATO}(\{h_l\})=-g_\mathcal{A}\sum_i \mathcal{A}_i-g_\mathcal{B}\sum_p \mathcal{B}_p-\sum_i h_la^z_l.
    \label{Hato}
\end{equation}
Here $\mathcal{A}_i=\prod_{i\in\partial l} a^x_l$ and $\mathcal{B}_p=\prod_{l\in p}a^z_l$ stabilize the clean toric-code ground state $\ket{\text{TC}}$, which satisfies $\mathcal{A}_i\ket{\text{TC}}=\mathcal{B}_p\ket{\text{TC}}=\ket{\text{TC}}$. $i, l, p$ label the sites, links, and plaquettes of the square lattice. $\{h_l\}$ are independent random variables drawn from a bimodal distribution $h_l=\pm h$ with equal probabilities $p=1/2$. 

Without disorders, the Hamiltonian possesses a magnetic 1-form symmetry generated by $U_m^{(1)}(\gamma)=\prod_{l\in\gamma} a^x_l$ and an electric 1-form symmetry $U_e^{(1)}(\tilde\gamma)=\prod_{l\in\tilde\gamma} a^z_l$, where $\tilde\gamma$ denotes a loop on the dual square lattice. The subscripts of the 1-form symmetry actions reflect the fact that the line operator of the $m$ anyon, the excitation associated with the plaquette violation $\mathcal{B}_p=-1$, is charged under $U_m^{(1)}(\gamma)$ if $\gamma$ encircles $p$. Similar, the line operator of the $e$ anyon, the excitation associated with a vertex violation $\mathcal{A}_i=-1$, is charged under $U_e^{(1)}(\gamma)$.

In the presence of weak disorders in the limit $h\ll g_{\mathcal{A},\mathcal{B}}$, the disordered Hamiltonian in Eq. \eqref{Hato} is in the TO phase, since the disorder strength never exceed the anyonic gap of $e$ and $m$. Therefore, we expect the disordered state can be smoothly connected to the clean limit toric code ground state $\ket{\text{TC}}$.

We now show that the ATO phase corresponds to the case where  $g_\mathcal{A}\ll h\ll g_\mathcal{B} $\cite{tsomokos2011interplay,1formswssb}. Since the disorder strength is within the gap of the $m$ anyon, $U_m^{(1)}$ is still an exact symmetry for $H_\text{ATO}$. Therefore, for each disorder realization $\{h_l\}$, the ground state $\ket{\{h_l\}}$ is still stabilized by $\mathcal{B}_p$ since disorder strength is within the gap of the $m$ anyons.  However, the stabilizer $\mathcal{A}_i$ will be randomly violated by the disorder. In other words, we have an ensemble of incoherently exited $e$ anyons due to disorder. We can write down a schematically ``fixed point" density matrix for the ATO (up to normalization):
\begin{equation}
\begin{aligned}
    \rho_\text{ATO}\propto &\sum_{s_i=\pm 1}\ket{\mathcal{A}_i=s_i,\mathcal{B}_p=1}\bra{\mathcal{A}_i=s_i,\mathcal{B}_p=1}\\
    =&\prod_p\frac{\mathbbm{1}+\mathcal{B}_p}{2}.
\end{aligned}
\label{rhoato}
\end{equation}
One can confirm that such a disordered state is an ATO with SWSSB of $U^{(1)}_m$ using the loop order parameter of $U^{(1)}_m$, $O_{\tilde\gamma}=\prod_{l\in\tilde\gamma}a^x_l$: 
\begin{equation}
\begin{aligned}
&U^{(1)}_m(\gamma)\rho_\text{ATO}=\rho_\text{ATO}U^{(1)}_m(\gamma)=\rho_\text{ATO}\\
&\Tr[\rho_\text{ATO}O_{\tilde\gamma}]=0,\ \frac{\Tr[\rho_\text{ATO}O_{\tilde\gamma}\rho_\text{ATO}O_{\tilde\gamma}]}{\Tr[\rho_\text{ATO}^2]}=1
\end{aligned}
\end{equation}
for any loop $\gamma$ and $\tilde\gamma$ on the original and dual lattice, respectively. The first line confirms that the $U^{(1)}_m$ is an exact symmetry of $\rho_\text{ATO}$ and the second line shows the expected behaviors of the correlations function for an ATO.\footnote{The fidelity loop expectation value $\langle O_{\tilde\gamma}\rangle_2^F$ is also 1.}

Another ATO with randomly excited $m$ anyons can be similarly constructed by disorders of $a^x$ instead of $a^z$ in $H_\text{ATO}$.

In the rest of the section, we will mainly focus on a special type of 2+1$D$ TO: the deconfined phase in the gauge theory with a finite Abelian gauge group, denoted by $N$, where the physical picture of the disordered toric code generalizes. The anyons in such a TO are point excitations that correspond to the gauge charge and gauge flux\footnote{For a discrete gauge theory in $(d+1)D$, the gauge flux excitation has spacial dimension $d-2$.}.  The gauge flux anyons form a fusion group $N$, and the gauge charges form a fusion group $\hat{N}=\text{Hom}[N,U(1)]$ that is dual to $N$. The discrete gauge theory possesses two sets of 1-form symmetries in the deconfined phase: the ``magnetic" 1-form symmetry generated by the Wilson loops, and the ``electric" 1-form symmetry generated by the 't Hooft loops, which are both spontaneously broken due to deconfinement. An ATO can be obtained from the clean deconfined phase by randomly exciting either gauge charges or gauge fluxes through disorder, while the other sector remains coherent. For example, if the disorder randomly excites the gauge charges, the electric 1-form symmetry is {\it explicitly} broken down to an average symmetry. 
Meanwhile, the magnetic 1-form symmetry is in a SWSSB state. That is because the symmetry generator of the electric 1-form is the order parameter of the magnetic 1-form symmetry.
Since the density matrix still possesses an average electric 1-form symmetry, the R\'enyi-2 expectation value of order parameter for the magnetic 1-form symmetry, $\langle O_\gamma\rangle_2$, is hence non-vanishing. The disordered toric code state is one example of such a scenario. Similarly, an ATO with randomly excited gauge fluxes can also be obtained in this way, which is an SWSSB state of the electric 1-form symmetry.

\subsection{Review of anomaly-compatible TO in the clean limit}
\label{sec:SET}
We now briefly review the construction of the topologically ordered (TO) state that are compatible with the exact anomaly of $G$ in $(d+1)D$ where $d\geq 2$. 
To construct such a gauge theory, we first need to extend the symmetry group $G$ to a larger group $G'$ via the short exact sequence
\begin{equation}
    1\to N\to G'\to G\to 1,
\end{equation}
where $N$ is a discrete Abelian group and a normal subgroup of $G'$: $G=G'/N$. Such a group extension is characterized by a map $\mathfrak{e}: G\times G\to N$ in the second cohomology group, $\mathfrak{e}\in {\cal H}^2(G,N)$. Refs. \onlinecite{wangwenwitten,tachikawa2020gauging,WenPotterGapless} show that if one is able to find a group extension to $G'$ through a finite group $N$ that trivializes the anomaly, the gauge theory with gauge group $N$ is then compatible with the $G$-anomaly. Here the trivialization of the anomaly means that the cohomology group $\mathcal{H}^{d+2}(G',U(1))$, is trivial, in contrast to the non-trivial element in $\mathcal{H}^{d+2}(G,U(1))$ that corresponds to $G$-anomaly. Furthermore, when $G$ is finite, such a group $N$ always exists. \footnote{We note that for some exact anomalies, there is an obstruction of having an anomaly-compatible topologically ordered state, see Refs  \onlinecite{gaiottotheta,garcia20178d,cordova2019anomaly,cheng2023gauging,yang2023gapped,kapustin2024anomalous}.}

After the anomaly is trivialized via the group extension to $G'$, the TO can be obtained by gauging the normal subgroup $N$, which brings the global symmetry back to $G$. The TO is the deconfined phase of the $N$ gauge theory, both the gauge charge and gauge flux anyons transform non-trivially under symmetry. The gauge charge of $N$ carries a projective representation of $G$, which is inherited from the structure of group extension $n=\mathfrak{e}(g_1,g_2)$. In addition, the anomaly also assigns a projective representation of $G$ to the gauge flux, which can be represented by an element in the second group cohomology ${\cal H}^2_\rho(G,\hat{N})$, where $\rho$ is the symmetry action of $G$ on the anyons in the $N$-gauge theory\cite{barkeshliset,chengexactlysolvable}.
A TO where its anyons transform non-trivially under global symmetry is called a symmetry-enriched TO. In general, if a TO carries the anomaly of $G$, the way the anyons transform under the global symmetry cannot be realized in an anomaly-free 2+1$D$ system\cite{barkeshliset,chensemion}.

For example, we consider symmetry-enriched TO states that are compatible with the 2+1$D$ exact anomalies $\Omega_{1,2}$ of $G=A\times K$ with $A=K=\Z_2$ discussed in \secref{sec:anomaly}. To trivialize the anomaly $\Omega_1$, we pick $N=\Z_2$ and extend the global symmetry group $G$ by \beqn\mathfrak{e}(ak^i,a^jk)=n,\label{frake}\eeqn where $a,k,n$ are the non-trivial group elements of $A,K,N$ and $i,j=0,1$ (all the other situations are mapped to the trivial group element in $N$). The extended global symmetry group becomes $G'=D_4$.\footnote{To be clear, $D_4$ stands for the dihedral group with 8 elements. Such a group is sometimes denoted as $D_8$.} We then gauge $N$ in $G'=D_4$, which yields a $\Z_2$ gauge theory enriched by $G=A\times K=\Z_2^2$. From $\mathfrak{e}$, the $e$ anyon carries a projective representation between $A$ and $K$, and $m$ anyon transforms as a half charge of $K$. We denote such a symmetry-enriched TO state as $e_{AK}m_K$, similar to the notation used in \refcite{chongwangbspt} where the subscripts label how the anyon transforms under $A\times K$. Alternatively, the anomaly can be trivialized via a group extension to $G'=A\times K'=\Z_2\times\Z_4$, in which case $\mathfrak{e}'(a^i k,a^j k)=n$. After gauging $N$, the global symmetry $G$ fractionalizes as $e_Km_{AK}$. These two symmetry-enriched TOs are essentially the same after relabelling the $e$ and $m$ anyons.

For the other anomaly $\Omega_2$, which is equivalent to $\Omega_1$ but with the roles of $A$ and $K$ interchanged, one can similarly extend the global symmetry to $G'=D_4$ or $G'=A'\times K=\Z_4\times\Z_2$ via $N=\Z_2$. Gauging $N$, we arrive at two equivalent symmetry enriched TOs $e_{AK}m_A$ and $e_Am_{AK}$.

\subsection{TO and ATO compatible with the AEM anomaly}
\label{sec:TOandATO}
We are now ready to discuss disordered TO and ATO enriched by $G=A\times K$ that are compatible with the AEM anomaly of $G$ in 2+1$D$. It is easy to see that if a clean TO enriched by the exact symmetry $G$ is compatible with the $G$-anomaly in the clean limit, weakly disordering it within the TO phase will certainly give a disordered state that is compatible with the AEM anomaly. Therefore, the most important question here is whether a strong enough disorder is capable of driving the TO into an ATO by randomly exciting some anyons in the TO without causing spontaneous symmetry breaking of either the exact symmetry $K$ or the average symmetry $A$. 

If an anyon $e$ in the TO transforms projectively under some subgroup $H$ of $G$, creating $e$ anyon excitations will break $H$. Therefore, in the clean limit, coherently proliferating such an anyon leads to either SSB of $H$ or some gapless degrees of freedom. In a disordered system where $e$ anyons are incoherently excited, $H$ can still be an average symmetry of the disordered states, since different symmetry-breaking states caused by $e$ anyon excitations can transform into each other under the action of $H$. Furthermore, if $H=A$, incoherently exciting $e$ will no longer cause any symmetry breaking now that $A$ is already an average global symmetry\cite{ASPT2}. Therefore, in the clean TO that carries the exact anomaly of $G=A\times K$, if there exists an anyon $e$ that transforms non-trivially only under the subgroup $A$, the AEM anomaly of $G=A\times K$ can instead be carried by an ATO where $e$ can be incoherently excited. In contrast, if all anyons in the TO transform non-trivially under $K$, we cannot have an anomaly-compatible ATO without SSB or SWSSB of the (0-form) symmetry $K$.

Now consider the 2+1$D$ discrete gauge theory constructed in Sec. \ref{sec:SET} that carries the exact anomaly. Since the charge and flux sectors are dual to each other, we mainly focus on the possible ATO with incoherently excited gauge charges. Recall that the projective representation carried by the gauge charge is inherited from the group extension $\mathfrak{e}$. Therefore, if the $G$-anomaly can be trivialized by a group extension that only involves $A$, the gauge charges will transform projectively only under $A$. Consequently, they can be randomly excited without breaking the average symmetry $A$ or the exact symmetry $K$ in a disordered system. Our argument can be summarized as the following: 
\begin{tcolorbox}
An ATO compatible with an AEM anomaly of $G=A\times K$ exists if there exists a group extension via a finite Abelian group $N$ of the form \beqn 1\to N\to G'=A'\times K\to G=A\times K\to 1\label{groupext}\eeqn that trivializes the exact anomaly of $G$.
\end{tcolorbox}
We expect that a similar result holds in higher dimensions, where a deconfined phase of the discrete gauge theory of $N$ with randomly excited gauge charges is compatible with the AEM anomaly if the group extension in Eq. \eqref{groupext} exists. We also note that such a condition is sufficient but not necessary for having an anomaly-compatible ATO. 

As an example, we discuss possible TO and ATO that are compatible with the two AEM anomalies of $G=A\times K=\Z_2\times\Z_2$ in 2+1$D$, which are obtaining from ``disordering" the exact anomalies $\Omega_{1,2}$ (which are introduced in \ref{sec:anomaly}). These two AEM anomalies will still be referred to as $\Omega_{1,2}$. Recall from Sec. \ref{sec:SET} that the exact anomaly $\Omega_1$ can be trivialized by extending to $G'=A\times K'=\Z_2\times \Z_4$, but not to a group $G'$ of the structure $A'\times K$. Therefore, it is impossible to construct an ATO via the discrete gauge theory construction mentioned above. In \secref{sec:cupprod}, we will use a field theory perspective to further argue that there is, in fact, no ATO compatible with the AEM anomaly of $\Omega_1$. Indeed, in the symmetry-enriched TO that carries the exact anomaly of $\Omega_1$, both the $e$ and $m$ anyons transform non-trivially under $K$, so none of them are allowed to be randomly excited without symmetry breaking. In contrast, the exact anomaly of $\Omega_2$ can be extended to $G'=A'\times K=\Z_4\times\Z_2$. Hence, there exists an ATO that is compatible with the AEM anomaly of $\Omega_2$. Indeed, in the symmetry-enriched TO $e_{A}m_{AK}$ that carries the exact anomaly of $\Omega_2$, the $e$ anyon now only transforms under the average symmetry $A$. Therefore, an ATO can be obtained by randomly exciting the $e$ anyons. We will confirm these results using a solvable lattice model in \secref{sec:z22model}.

\begin{figure}
    \centering
    \includegraphics[width=4.5cm]{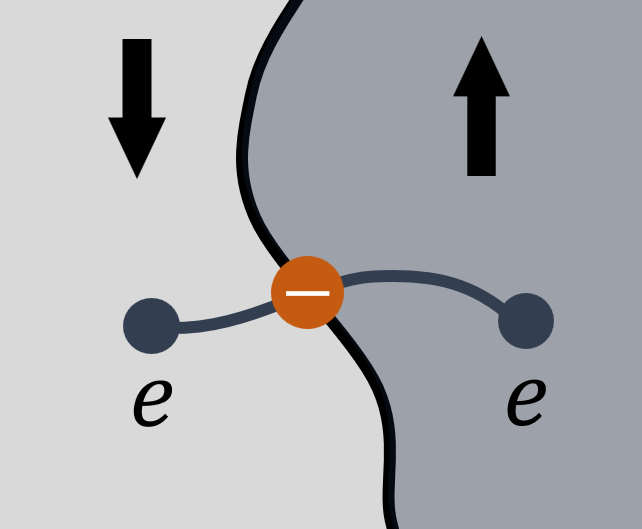}
    \caption{In the $e_{AK}m_K$ state, a pair of $e$ anyons created by an open Wilson line is charged under $K$ when the line intersects with an $A$ domain wall.}
    \label{fig:dwtunnel}
\end{figure}

A useful way to understand the obstruction of having an ATO that is compatible with the AEM anomaly of $\Omega_1$ is to consider an open Wilson line in the $N=\Z_2$-gauge theory that excites a pair of $e$ anyons in the $e_{AK}m_K$ state in a typical disordered sample where $A$-spins are pinned by the random-field disorder (see \figref{fig:dwtunnel}). Since $e$ carries a projective representation between $A$ and $K$, when the open Wilson line crosses a domain wall of $A$, it creates a local $K$-charge (which will be clarified further in the subsequent subsections). 
 In the entire disorder ensemble of $A$, domain walls appear randomly in each disorder realization. Thus, anyonic excitations will introduce random $K$ charges to each sample depending on the number of $A$ domain walls that cross the $e$ anyon lines modulo $2$. As such, the exact symmetry $K$ is broken down to an average symmetry if one allows random $e$ anyon excitations. A similar argument in the clean limit implies that proliferating the $e$ anyon will induce SSB of $K$ (to a trivial group).

We note that for an ATO, one cannot solely use the symmetry fractionalization pattern of the anyons that have not been randomly disordered in determining whether the ATO, by itself, is anomalous or not. The distinction needs to be exposed by creating a coherent anyonic excitation on top of the disordered state. Similarly, one cannot simply tell if a fully ordered SSB state is anomalous without considering the effect of domain-wall excitations.
As an example, the AEM anomaly of $\Omega_2$ can be carried by the ATO obtained from incoherently exciting the $e$ anyons in the $e_Am_{AK}$ state. The only symmetry fractionalization in the ATO that remains meaningful is $m_{AK}$. Such a fractionalization pattern is also realizable by an ATO derived from a TO enriched by anomaly-free global symmetries, in which the $e$ anyon transforms as the linear representation of both $A$ and $K$ and the $m$ anyon carries the projective representation between $A$ and $K$ (see Ref. \onlinecite{lu2016classification} and also \appref{app:VW}).

\begin{figure}
    \centering
    \includegraphics[width=4cm]{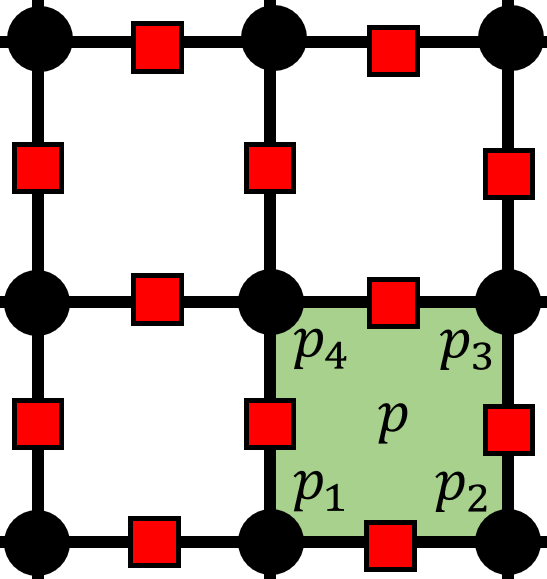}
    \caption{A lattice realization of the 2+1$D$ anomaly of $G=\Z_2^2$. Here, each lattice site (black dot) hosts two flavors of Pauli spins denoted by $\sigma$ and $\tau$. Each plaquette is labeled by $p$, whose four corners are denoted by $p_{1,2,3,4}$ in counterclockwise order. To extend symmetry to $D_4$ by $N$, we further introduce $\mu$ spins on each lattice site. The $\Z_2$ gauge field of $N$ lives on bonds between lattice sites marked by red squares.}
    \label{fig:z22lattice}
\end{figure}

\subsection{Solvable model for symmetry-enriched TO and ATO}
\label{sec:z22model}
To bolster our arguments above, we explicitly construct an exactly solvable lattice model of the clean limit symmetry-enriched TO $e_{AK}m_K$ that carries the exact anomaly of $\Omega_1$ of $G=A\times K=\Z_2^2$ in 2+1$D$, and discuss its fate against various forms of disorders that can potentially drive it towards the ATO state. To discuss the AEM anomaly $\Omega_2$ in the same lattice representation, \textit{in this subsection}, we redefine the AEM anomaly of $\Omega_2$ as the disordered version of the exact anomaly of $\Omega_1$ but with the subgroups $A$ being \textit{exact} and $K$ being \textit{average}. This is equivalent to the original definition of the AEM anomaly of $\Omega_2$.

The anomalous symmetry actions of $A$ and $K$ for the exact anomaly $\Omega_1$ can be represented on a square lattice with two flavors of Pauli spins $\sigma_i$ and $\tau_i$ on the lattice sites:
\begin{equation}
    U_A=\prod_i \sigma^x_i,\ U_K=\prod_p C(-\tau^z_{p_1}\tau^z_{p_3},\tau^z_{p_2}\tau^z_{p_4},\sigma^z_{p_3})\prod_i\tau^x_i,
    \label{UAUK2}
\end{equation}
where $C(A,B,C)=e^{\frac{\ii\pi}{8}(1-A)(1-B)(1-C)}$ is the controlled-control operator between three Pauli operators $\rho,\sigma$ and $\tau$. $p_{1,2,3,4}$ denote the four lattice sites that forms the plaquette $p$ in the counterclockwise order, see \figref{fig:z22lattice}. These symmetry representations can be directly obtained from group cohomology, see \appref{app:cohomology}.

Now, we begin the construction of the symmetry-enriched TO in the clean limit.Recall from Sec. \ref{sec:SET} that the exact anomaly $\Omega_1$ can be trivialized by extending the global symmetry to $G'=D_4$ or $\Z_2\times \Z_4$ via group extension by $N=\Z_2$. To gauge $N$ and obtain the anomaly-compatible TO, we introduce the gauge field of $N$ on the links $l$ of the square lattice (see Fig. \ref{fig:z22lattice}), whose Pauli operators are denoted by $a^{x,z}_l$.  As shown in \appref{app:VW}, after symmetry extension and gauging, the wave function of the symmetry-enriched TO  can take the following form: the $N$-gauge theory, whose wave function is denoted by $\ket{a}$, is coupled to the ``matter fields" $\sigma$ and $\tau$ via a set of unitary operators\beqn\ket{\psi}=\mathcal{V}\mathcal{W}\left(\ket{\sigma,\tau}\otimes\ket{a}\right),\label{psischematic}\eeqn where $\ket{\sigma,\tau}$ represents a possible state for the $\sigma$ and $\tau$ spins. As far as the symmetry action is concerned, the unitaries $\mathcal{V}$ and $\mathcal{W}$ describe the coupling between the $\Z_2$ gauge field and the ``matter fields" $\sigma$ and $\tau$, which gives rise to the projective representation carried by the gauge charge and flux. Their explicit forms are \beqn\mathcal{V}=\prod_pC(\tau^z_{p_1},\tau^z_{p_1}\tau^z_{p_2},a^z_{p_2p_3})C(\tau^z_{p_1},\tau^z_{p_1}\tau^z_{p_4},a^z_{p_4p_3}).\label{calV}\eeqn
and 
\beqn\mathcal{W}=\prod_{l} C\left( a^x_l,\sigma^z_{l_1}\sigma^z_{l_2},\tau^z_{l_1}\right).\label{calW}\eeqn Here, $l_{1,2}$ are two sites of the bond $l$, and $l_1$ is always below or to the left of $l_2$. $p_i p_j$ means the edge connecting the two sites $p_i$ and $p_j$. The derivations of $\mathcal{V}$ and $\mathcal{W}$ can be found it App. \ref{app:VW}.

In the $\mathcal{V}\mathcal{W}$-transformed frame, the global symmetries are 
\begin{equation}
\begin{aligned}
    U_A'&\equiv\mathcal{VW} U_A \mathcal{W}^{-1} \mathcal{V}^{-1}=\prod_i\sigma^x_i,\\  
    U_K'&\equiv\mathcal{VW} U_K \mathcal{W}^{-1}\mathcal{V}^{-1}\\
    &=\prod_p C\left(\prod_{l\in \partial p}a^z_{l},\tau^z_{p_1}\right)
    \prod_i C\left(\prod_{i\in\partial l}
        a^x_{l},\sigma^z_i\right)\tau^x_i.
\end{aligned}
\label{uprime}
\end{equation}
These symmetry representations capture the fractionalization pattern $e_{AK}m_K$ of the global symmetries. The $N$ charge (the $e$ anyon), which corresponds to a local violation of the Gauss law at the site $i$ by $\mathcal{A}_i=\prod_{i\in\partial l}a^x_{l}=-1$, leads to a local projective representation between $A$ and $K$ represented by $U'_A|_i=\sigma^x_i$ and $U'_K|_i=\sigma^z_i\tau^x_i$. A $\pi$-flux of $N$ (the $m$ anyon) at the plaquette $p$ with $\mathcal{B}_p=\prod_{l\in \partial p}a^z_{l}=-1$ 
carries the half charge of $K$ as indicated by $U'_k|_{p_1}=\tau^z_{p_1}\tau^x_{p_1}=\ii\tau^y_{p_1}$. Furthermore, the fact that the $e$ anyon line is charged under $K$ when crossing a domain wall of $A$ can now be explicitly shown using the symmetry action. Assigning $\sigma^z=1$ on one side of the domain wall and $\sigma^z=-1$ on the other side, the non-on-site part of the symmetry action of $K$ in \eqnref{uprime} now simply acts on the domain wall:
\begin{equation}
    \prod_i C\left(\prod_{i\in\partial l}
        a^x_{l},\sigma^z_i\right)\overset{DW_\sigma}{\longrightarrow}\prod_{l\in DW_\sigma}a^x_l.
\end{equation}
Meanwhile, the open Wilson line that creates a pair of $e$ anyons can be represented by $W(\gamma)=\prod_{l\in\gamma}a^z_l$. Therefore, we have
\begin{equation}
    U_K'W(\gamma)\left(U_K'\right)^{-1}=(-1)^{|\gamma\cap DW_\sigma|}W(\gamma),
    \label{dwcharge}
\end{equation}
where $|\gamma\cap DW_\sigma|$ represents the number of times the Wilson line $\gamma$ intersects with $DW_\sigma$.

Since the anomaly-compatible TO under discussion is the deconfined phase of the $N$ gauge theory, the gauge sector of the state is in the state $\ket{a}=\ket{\text{TC}}$, so that $\mathcal{A}_i\ket{a}=\mathcal{B}_p\ket{a}=\ket{a}$. The wave function of the entire system now reads
\begin{equation}
    \ket{\psi_{e_{AK}m_{K}}}=\mathcal{VW}\left(\ket{+}_\sigma\otimes\ket{+}_\tau\otimes \ket{\text{TC}}\right),
\end{equation}
where we have taken $\ket{+}_{\sigma}=\bigoplus_{i}\ket{\sigma^x_i=1}$ and similarly for $\ket{+}_\tau$ to ensure the $A\times K$ symmetry of the system.  To set the stage for the discussion of the effect of disorders, we write down an exactly solvable parent Hamiltonian for $\ket{\psi_{e_{AK}m_{K}}}$:
\begin{equation}
\begin{aligned}
    H_{e_{AK}m_{K}}=&-\mathcal{VW}\left[\sum_i \sigma^x_i\left(1+\mathcal{A}_i\right)+\sum_p\tau^x_{p_1}\left(1+\mathcal{B}_{p}\right)\right.\\
    &\left. +g_\mathcal{A}\sum_i\mathcal{A}_i+g_\mathcal{B}\sum_p\mathcal{B}_p\right]\mathcal{W}^{-1}\mathcal{V}^{-1}.
\end{aligned}
\label{HSET}
\end{equation}

We now discuss possible disorders that can drive the system towards the ATO. For convenience, we will work in the $\mathcal{VW}$-transformed frame in which symmetries are represented by $U_{A,K}’$. We first consider $\Omega_2$, since we have argued for the possibility of an ATO by randomly exciting the $m$ anyon in the $e_{AK}m_K$ state (as a reminder, $K$ is the average symmetry for $\Omega_2$). From the clean limit Hamiltonian in \eqnref{HSET}, we introduce the following disorders that randomly excite the $m$ anyon and explicitly break $K$ down to an average symmetry:
\begin{equation}
\begin{aligned}
    H_{\Omega_2}=&H_{e_{AK}m_{K}}-\sum_l h^{(a)}_l a^x_l+\sum_i h_i \tau^z_i\\
    \approx& -\mathcal{VW}\left\{\sum_l h^a_l a^x_l C(\tau^z_{i_1(l)},\tau^z_{i_2(l)})+\sum_i \left[h_i \tau^z_i\right.\right.\\
    &\left.\left.+ \sigma^x_i\left(1+\mathcal{A}_i\right)+g_\mathcal{A}\mathcal{A}_i\right]+g_\mathcal{B}\sum_p\mathcal{B}_p\right\}\mathcal{W}^{-1}\mathcal{V}^{-1},
\end{aligned}
\label{HGK}
\end{equation}
where $h^{(a)}_l=\pm h^{(a)}$ and $h_i=\pm h$ both with probabilities $\frac{1}{2}$ and in the second line we have taken the strong disorder limit $h^{(a)},h\gg 1$ and $g_\mathcal{A}>h^{(a)}>g_\mathcal{B}$. $\tau^z_{i_{1,2}(l)}$ are the two spins entangled with $a^z$ through $\mathcal{V}$. In the second line, we have omitted the $\tau^x_{p_1}\left(1+\mathcal{B}_{p}\right)$ term in $H_{\Omega_2}$ since it does not commute with the strong $\tau^z$ disorder terms. Because the original symmetries $U_{A,K}$ do not involve the gauge field, the random $a^x$ disorder is always symmetric. In the $\mathcal{VW}$-transformed frame, the coefficient of the random $a^x$ disorder, $h^a_l C(\tau^z_{i_1(l)},\tau^z_{i_2(l)})$, is still a random number $\pm h^{(a)}$ with probabilities $\frac{1}{2}$ for
any quenched values of $\tau^z$. As such, the gauge sector will be been driven to an ATO whose fixed-point density matrix is \beqn \rho_\text{ATO}'=\prod_i\frac{\mathbbm{1}_a+\mathcal{A}_i}{2}.\eeqn Collecting the ground states for every disorder realization, the density matrix of the ensemble of ground states can be written as
\begin{equation}
    \rho_{\Omega_2}=\mathcal{VW}\left(\ket{+}\bra{+}_\sigma\otimes\frac{\mathbbm{1}_\tau}{2^{N_\tau}}\otimes\rho_\text{ATO}'\right)\mathcal{W}^{-1}\mathcal{V}^{-1},
\end{equation}
which has an exact $A$ symmetry and average $K$ symmetry (without SSB or SWSSB). 

Next, we move on to the AEM anomaly of $\Omega_1$. As argued in \secref{sec:TOandATO}, it is impossible to obtain an anomaly-compatible ATO for the AEM anomaly of $\Omega_1$ by randomly exciting either the $e$ or $m$ anyon, since both of them transforms non-trivially under $K$. Randomly exciting any of them induces SWSSB of the exact symmetry $K$. To see this in the lattice model, we construct the following Hamiltonian where $e$ is randomly excited by a random field disorder of $a^z$ and $A$ is broken down to an average symmetry ($h_l^{(a)}$ and $h_i$ follows the same distribution as in \eqnref{HGK}):
\begin{equation}
\begin{aligned}
    H_{\Omega_1}=&H_{e_{AK}m_{K}}-\sum_l h^{(a)}_l a^z_l+\sum_i h_i \sigma^z_i\\
    \approx& -\mathcal{VW}\left\{\sum_l h^{(a)}_l a^z_l C(\sigma^z_{l_1}\sigma^z_{l_2},\tau^z_{l_1})\right.\\
    &\left.+\sum_i \left(h_i \sigma^z_i+g_\mathcal{A}\mathcal{A}_i\right)\right.\\
    &\left.+\sum_p\left[g_\mathcal{B}\mathcal{B}_p+\tau^x_{p_1}\left(1+\mathcal{B}_{p}\right)\right]\right\}\mathcal{W}^{-1}\mathcal{V}^{-1}.
\end{aligned}
\end{equation}
We again consider the strong disorder limit $g_\mathcal{B}> h^{(a)}>g_\mathcal{A}$ and $h,h^{(a)}\gg 1$, so that the term $\sigma^x_i\left(1+\mathcal{A}_i\right)$ in $H_{e_{AK}m_K}$ can be omitted. We notice that the random $a^z$ disorder now depends on whether these is a $\sigma^z$ domain wall located between $l_1$ and $l_2$. Away from the domain wall, the $a^z$ disorder drives the gauge sector to an ATO where the density matrix is schematically $\prod_p\frac{\mathbbm{1}+\mathcal{B}_p}{2}$. However, on the $\sigma^z$ domain wall, the term becomes $h_l^{(a)}a^z_l\tau^z_{l_1}$. Since $h^{(a)}\gg 1$, such an interaction on the $A$ domain wall aligns (or anti-aligns) $\tau^z_{l_1}$ with $a^z_l$ on the $\sigma^z$ domain wall in the $\mathcal{VW}$-transformed frame whenever $l\in DW_\sigma$. In fact, this is a direct consequence of the fact that the $e$ anyon line is charged under $K$ when it intersects an $A$ domain wall (see \figref{fig:dwtunnel} and also \eqnref{dwcharge}), which in this case is created by the $a^z$ disorders.

To confirm the SWSSB of $K$ after the $e$ anyons in such a disordered state, we calculate the R\'enyi-2 $K$ charge correlator using the density matrix $\rho_{\Omega_1}$ of the disordered ground states of $H_{\Omega_1}$:
\begin{equation}
\begin{aligned}
    &\frac{\Tr[\rho_{\Omega_1}\tau^z_i\tau^z_j\rho_{\Omega_1}\tau^z_i\tau^z_j]}{\Tr[\rho_{\Omega_1}^2]}=\frac{\Tr[\rho_{\Omega_1}^\mathcal{VW}\tau^z_{i}\tau^z_{j}\rho_{\Omega_1}^\mathcal{VW}\tau^z_{i}\tau^z_{j}]}{\Tr[\rho_{\Omega_1}^2]}\\
    &\approx\left(\frac{1}{2}\right)^2\frac{\Tr[\rho_{\Omega_1}^\mathcal{VW}a^z_{l_i}a^z_{l_j}\rho_{\Omega_1}^\mathcal{VW}a^z_{l_i}a^z_{l_j}]}{\Tr[\rho_{\Omega_1}^2]} \approx \frac{1}{4},
\end{aligned}
\end{equation}
where $\rho_{\Omega_1}^\mathcal{VW}=\mathcal{WV}\rho_{\Omega_1}\mathcal{V}^{-1}\mathcal{W}^{-1}$ and $l_{i,j}$ are the bonds on the $\sigma^z$ domain wall near the sites $i$ and $j$. The first equality in the second line approximately holds since each site $i$ has probability $\frac{1}{2}$ of being on the domain wall of $\sigma^z$, in which case $\tau^z_{i}$ will be energetically locked to either $a^z_{l_i}$ or $-a^z_{l_i}$ depending on the sign of $h^{(a)}_l$. Away from the domain wall, the $\tau$ spins are polarized in $\tau^x=1$ by the term $\tau^x_{p_1}\left(1+\mathcal{B}_{p}\right)$ in the Hamiltonian $H_{\Omega_1}$. Therefore, they do not contribute to the R\'enyi-2 correlator. In the second line, we use the fact that $a^z_{l_i}a^z_{l_j}$ commutes the density matrix of the ATO. 

The disordered lattice model with random $m$ anyon excitations can be similarly constructed, and such a model in the strong disorder limit will also exhibit SWSSB of $K$.

\subsection{Field-theoretical arguments: anomaly inflow in doubled Hilbert space}
\label{sec:cupprod}
The possibility of an anomaly-compatible ATO enriched by the global symmetries can be discussed using an field-theoretical approach via coupling the disordered system to background gauge fields of the global symmetries.

In the clean limit, the mixed anomaly $\Omega_1$ can be probed by background $\Z_2$ gauge fields of $A$ and $K$. Since the global symmetry $G=A\times K$ is anomalous, the partition function of the 2+1$D$ system is only gauge-invariant when the system lives on the boundary of a 3+1$D$ bulk $M_4$, whose response to the gauge fields is:
\begin{equation}
    Z_{\Omega_1}\sim\exp\left(\ii\pi\int_{M_4} \mathfrak{a}\cup\mathfrak{k}\cup\mathfrak{k}\cup\mathfrak{k}\right),
    \label{ak^3}
\end{equation}
where $\mathfrak{a},\mathfrak{k}\in H^1(M_4,\Z_2)$ are $\Z_2$ background gauge fields of $A$ and $K$, respectively.  The action on the exponent of \eqnref{ak^3}, called the anomaly inflow, is the response of the 3+1$D$ SPT state to external $A$ and $K$ gauge fields (after the gapped matter field of the SPT state is integrated out). To see this, we show that the action matches the decorated defect construction: suppose the action lives in a cartesian product space $M_4=S_1\times M_3$. An domain wall of $A$ can be created by threading a ``$\pi$-flux" of $A$ through $S_1$, which is achieved by modifying the background gauge field $\mathfrak{a}\to\mathfrak{a}+\Delta\mathfrak{a}$, such that $\int_{S_1}\Delta\mathfrak{a}=1$. Such a modification generates a new factor $\exp\left(\ii\pi\int_{M_3}\mathfrak{k}\cup\mathfrak{k}\cup\mathfrak{k}\right)$, which describes a 2+1$D$ Levin-Gu state probed by the $\Z_2$ gauge field $\mathfrak{k}$\cite{kapustin2014anomalies,wen2015construction, wang2015field}. Similarly, the inflow of the exact anomaly $\Omega_2$ can be described by the following partition function
\begin{equation}
    Z_{\Omega_2}\sim\exp\left(\ii\pi\int_{M_4} \mathfrak{a}\cup\mathfrak{a}\cup\mathfrak{a}\cup\mathfrak{k}\right).
    \label{a^3k}
\end{equation}

Meanwhile, the TO can be viewed as an SSB state that is compatible with the mixed anomaly in 2+1$D$ between a pair of 1-form $\Z_2$ symmetries $\Z_{2e}^{(1)}\times\Z_{2m}^{(1)}$ represented by $U^{(1)}_{e,m}$. This 1-form anomaly is captured by the action in the 3+1$D$ bulk $M_4$ as
\begin{equation}
    \exp\left(\ii\pi\int_{M_4}\mathfrak{b}_e\cup\mathfrak{b}_m\right),
    \label{bebm}
\end{equation}
where $\mathfrak{b}_{e,m}\in H^2(M_4,\Z_2)$ are 2-form background gauge fields of $\Z_{2e,m}^{(1)}$, respectively. In the $e_{AK}m_K$ state, the $e$ anyon carries a mutual projective representation between $A$ and $K$, thus the background gauge field of $\Z_{2e}^{(1)}$ can be identified with the background gauge fields of $A$ and $K$ as $\mathfrak{b}_e= \mathfrak{a}\cup\mathfrak{k}$. Similarly, $\mathfrak{b}_m=\mathfrak{k}\cup\mathfrak{k}$.\cite{benini20192, hsin2021discrete} We then see that the two anomaly inflows in \eqnref{ak^3} and \eqnref{bebm} agrees with each other. In the lattice model, these identifications are made explicit through the unitaries $\mathcal{V}$ and $\mathcal{W}$ in Eqs. \eqref{calV} and \eqref{calW}. 

We now use the field-theoretical formalism to discuss TO and ATO states that are compatible with the AEM anomalies $\Omega_{1,2}$. To write down the anomaly inflow of the AEM anomaly, we utilize the doubled Hilbert space formalism by mapping the density matrix $\rho=\sum_\alpha p_\alpha\ket{\psi_\alpha}\bra{\psi_\alpha}$ of the system to a quantum state $\ket{\rho}\rangle=\sum_\alpha p_\alpha\ket{\psi_\alpha}_L\otimes\ket{\psi_\alpha}_R$ through the Choi–Jamiołkowski isomorphism\cite{jamiolkowski1972linear,choi1975completely}. The Hilbert space is then doubled, and so does the exact symmetry $K$, which now becomes a pair of independent symmetries $K_L$ and $K_R$, which acts only on the left and right subspace of the doubled Hilbert space (the action of $K_R$ needs to be complex conjugated). This means that the background gauge field for $K_{L,R}$, $\mathfrak{k}_{L,R}$, can also be independently introduced. In order for the doubled state to correspond to a physical density matrix (that is Hermitian), it should automatically be symmetric under the $\Z_2^*$ anti-unitary symmetry that swaps the left and right Hilbert spaces with complex conjugation. Since the left/right symmetries are also permuted by $\Z_2^*$, the exact symmetry $K$ after the Choi–Jamiołkowski isomorphism becomes
\begin{equation}
    (K_L\times K_R)\rtimes\Z_2^*.
\end{equation} Therefore, the response to the external gauge field $\mathfrak{k}_{L,R}$ should be  symmetric under $\Z_2^*:\mathfrak{k}_L\leftrightarrow\mathfrak{k}_R$. Meanwhile, the average symmetry $A$ acts simultaneously on both subspaces, i.e. there only exists a single diagonal $A$ gauge field $\mathfrak{a}$. From these symmetry perspective, the anomaly inflows of the AEM anomalies of $\Omega_{1,2}$ in the doubled Hilbert space can be written as (the superscript $\text{D}$ denotes that the inflow is in doubled Hilbert space)
\begin{equation}
    Z_{\Omega_1}^\text{D}\sim \exp\left[\ii\pi\int_{M_{4}}\left( \mathfrak{a}\cup\mathfrak{k}_L\cup\mathfrak{k}_L\cup\mathfrak{k}_L-\mathfrak{a}\cup\mathfrak{k}_R\cup\mathfrak{k}_R\cup\mathfrak{k}_R\right)\right]
    \label{ak^3double}
\end{equation}
and 
\begin{equation}
    Z_{\Omega_2}^\text{D}\sim \exp\left[\ii\pi\int_{M_{4}}\left( \mathfrak{a}\cup\mathfrak{a}\cup\mathfrak{a}\cup\mathfrak{k}_L-\mathfrak{a}\cup\mathfrak{a}\cup\mathfrak{a}\cup\mathfrak{k}_R\right)\right].
    \label{a^3kdouble}
\end{equation}

We now discuss the compatibility between the ATO and the AEM anomaly using such a formalism. The ATO with incoherently excited $e$ anyons is an SSB state compatible with the AEM anomaly between an average 1-form symmetry $\Z_{2e}^{(1)}$ and exact 1-form symmetry $\Z_{2m}^{(1)}$, whose anomaly inflow in the doubled Hilbert space can be written as
\begin{equation}
    Z_\text{ATO}^\text{D}\sim \exp\left[\ii\pi\int_{M_{4}}\left(\mathfrak{b}_e\cup\mathfrak{b}_{m,L}-\mathfrak{b}_e\cup\mathfrak{b}_{m,R}\right)\right].
    \label{bebmdouble}
\end{equation}
Comparing \eqnref{bebmdouble} with \eqnref{ak^3double} and \eqnref{a^3kdouble}, we see that the AEM anomaly inflow of $\Omega_2$ can be matched by the inflow of ATO if one assigns $\mathfrak{b}_e=\mathfrak{a}\cup\mathfrak{a}$ and $\mathfrak{b}_{m,L/R}=\mathfrak{a}\cup\mathfrak{k}_{L/R}$ in \eqnref{bebmdouble}. This confirms that the AEM anomaly of $\Omega_2$ can be carried by an ATO, where $e$ is randomly excited and $m$ still carries a projective representation of $A$ and $K$. In contrast, the inflow of the AEM anomaly of $\Omega_1$ in \eqnref{ak^3double} cannot be matched by \eqnref{bebmdouble}, which indicates the impossibility of having an anomaly-compatible ATO state.

We expect that the field-theoretical approach using the anomaly inflow in doubled Hilbert space can be easily generalized to higher dimensional AEM anomalies.

The possibility of an SWSSB state of $K$ to be compatible with $\Omega_{1,2}$ can also be understood using a similar AEM anomaly inflow argument starting from \eqnref{ak^3double} and \eqref{a^3kdouble}. In the doubled Hilbert space, a $K$-SWSSB state corresponds to an SSB state where the two symmetries $K_{L,R}$ are spontaneously broken down to a diagonal symmetry that transforms the $K$-spins in the two Hilbert spaces simultaneously. Therefore, the response of a $K$-SWSSB state to the external gauge fields $\mathfrak{k}_{L,R}$ should schematically be
\begin{equation}
    Z_\text{$K$-SWSSB}^\text{D}\sim\delta(\mathfrak{k}_L-\mathfrak{k}_R).
\end{equation}
The physical picture of the partition function is that there should be no response to the fluctuation that corresponds to the difference between the two Hilbert spaces. Consequently, the AEM anomalies will have a trivial inflow to the bulk, because Eqs. \eqref{ak^3double} and \eqref{a^3kdouble} vanishes after imposing the consequence of the $K$-SWSSB $\mathfrak{k}_L = \mathfrak{k}_R$. Therefore, the SWSSB state of $K$ is compatible with both AEM anomalies in Eqs. \eqref{ak^3double} and \eqref{a^3kdouble}.


\section{Conclusion and outlook}


In this work, we have explored various disordered phases that are compatible with the AEM anomaly of $G=A\times K$. We utilized a decorated-defect picture in the strong disorder limit that allowed us to physically understand the interplay between the constraint imposed by the quantum anomaly of the remaining exact symmetry and the localization effect of the disorder. With loosened constraints on global symmetry (with a part it being average symmetry) and quantum coherence, we found phases that have no clean limit analogs. One example is the glassy state with SWSSB, which is physically argued as a direct consequence of the lack of defect percolation of the average symmetry. We further constructed exactly solvable lattice models for SWSSB state that is compatible with a 1+1$D$ mixed anomaly.

We then discussed another possible AEM anomaly-compatible disordered phase intrinsic to mixed states, the ATO. In contrast to the SWSSB state, the ATO is not always compatible with the AEM anomaly. We argued for a sufficient condition for an AEM anomaly to be carried by an ATO, motivated by the construction of the anomaly-compatible TO in the clean limit. Namely, if the anomaly of the global symmetry group $G=A\times K$ can be trivialized by extending to a larger global symmetry group of the form $G'=A'\times K$ by a discrete group $N$, the $N$-charge in the topologically ordered $N$-gauge theory can be freely exited without breaking either the average symmetry $A$ or the exact symmetry $K$. This leads to an option for an anomaly-compatible ATO, i.e. an ensemble of random $N$-gauge charge excitations. On the contrary, if such an extension is impossible, exciting an $N$-charge induces SWSSB of the exact symmetry $K$, and hence, an anomaly-compatible symmetric ATO is impossible to construct in this way. Utilizing the Choi-Jamiołkowski map of the density matrix, we further bolstered our results using a field-theoretical argument of anomaly inflow in doubled Hilbert space. The cases where ATO is possible correspond to when the inflow of the AEM anomaly between the exact and average 1-form symmetries carried by the ATO can be matched to the AEM anomaly of 0-form global symmetry.

We conclude with some open questions and comments.
\begin{itemize}

\item \textit{Anomaly of a purely exact symmetry in disordered systems} - Throughout this work, we have mainly focused on disordered systems whose global symmetry has a direct product structure between exact and average symmetry, and the mixed anomaly is always between these two subgroups. It is possible that all the disorder realizations preserve the entire global symmetry group, in which case we have a disordered state that is compatible with the \textit{exact} anomaly. Such a scenario is recently discussed in Ref. \onlinecite{lessa2024mixed}, which proves that the exact anomaly constrains the multi-partite separability of the mixed state. However, this work does not emphasize on the possible physical feature of the mixed state that are compatible with the exact anomaly.

Gathering intuitions from our work, an SWSSB state of the \textit{entire} global symmetry should be an option for a mixed state compatible with the exact anomaly. Once the exact symmetry is spontaneously broken down to an average symmetry, the anomaly imposes no further constraint (recall from \secref{sec:disorderfeature} that an anomaly of an entirely average symmetry is trivial). For example, a $G$-symmetric infinite temperature state can always be compatible the exact anomaly of $G$, whose density matrix can be mathematically constructed by projecting the infinite temperature state to a fixed charge sector of $G$. Denoting such a projecting operator as $P_G$, the density matrix is simply proportional to it: $\rho\propto P_G$. Since the infinite temperature state always explicitly breaks the entire symmetry group $G$ down to an average symmetry, the projected density matrix will likely represent an SWSSB state of $G$.

Meanwhile, based on our arguments in Sec. \ref{sec:cupprod}, it seems impossible that any ATO without SWSSB can be compatible with an anomaly of a purely exact symmetry.

\item\textit{Anomaly of non-trivially extended global symmetries} - Another scenario we have not yet covered is when the global symmetry is not a simple direct product of the average and the exact symmetry. One special case of this is when the global symmetry group is mathematically a central extension of the average symmetry $A$ by the exact symmetry $K$. Ref. \onlinecite{ASPT2} has shown that such a global symmetry setting can give rise to intrinsically average SPT states, which have no clean limit analog. Consequently, the boundary of them would host a corresponding AEM anomaly of such a global symmetry. A similar situation is a mixed anomaly of a global symmetry that is non-trivially extended from an average spacial symmetry by an exact internal symmetry\cite{thorngren2018gauging,else2019crystalline}.
The physical consequence of these anomalies remains to be explored. 

\item \textit{Disordered phase transitions between anomaly-compatible phases} - Throughout the work, we mainly focused on the strong-disorder regime in arguing for the existence of various anomaly-compatible phases that are intrinsic to mixed states. Meanwhile, the nature of the phase transition between two disordered anomaly-compatible phases remains unclear. Since the phases we find in this work are all symmetry-breaking in some sense (with SSB or SWSSB of 0-form or higher-form symmetries), direct continuous phase transitions between them would be phase transitions between two different symmetry-breaking orders. For example, we showed in \secref{sec:stwmodel} that, both an SSB state of $A$ and an SWSSB state of $K$ can be realized using a single disordered Hamiltonian. If such a transition is continuous, it opens up the possibility of a disordered generalization of the deconfined quantum critical point (DQCP)\cite{senthilDQCP}, a transition between two different 0-form symmetry-breaking phases.

\item \textit{Entanglement constraint versus physical feature of the AEM anomaly} - Another important open direction is to connect various disordered AEM-anomaly-compatible phases with the quantum-information-theoretic constraints imposed by AEM anomaly, as proposed by Refs. \onlinecite{kimchirandommagnet,lessa2024mixed,wang2024anomaly}. In light of Ref. \onlinecite{kimchirandommagnet}, it would be meaningful to construct various anomaly-compatible disordered mixed states on the level of quantum states using symmetric quantum channels with the starting point being a trivial product state, in addition to their physical origin as ground states of disordered Hamiltonians. Such a construction should help understand better the quantum-information-theoretic constraint imposed by the AEM anomaly and their relation to physical observables. 

\end{itemize}

\textit{Acknowledgements} - YX and CMJ thank Yimu Bao, Zhen Bi, Meng Cheng, Tarun Grover, Jun Ho Son and Yizhi You for helpful discussions. YX thanks Zhen Bi, Zhu-Xi Luo, Cenke Xu, Carolyn Zhang, and Jian-Hao Zhang for collaboration on a related topic.
YX acknowledges support by the NSF through the grant OAC-2118310.
CMJ is supported by
Alfred P. Sloan Foundation through a Sloan Research Fellowship. This research was supported in part by grant NSF PHY-2309135 to the Kavli Institute for Theoretical Physics (KITP), where part of this work was performed.

\appendix
\onecolumngrid
\section{Anomalous symmetry action from group cohomology}
\label{app:cohomology}
In this appendix we will obtain the anomalous form of symmetry action directly from group cohomology that classifies the anomaly and also the corresponding SPT state in one higher dimension. SPT states protected by $G$ in $D$ space-time dimensions is classified by the cohomology group ${\cal H}^{D}[G,U(1)]$\cite{chencohomology}. The elements in the cohomology group are represented by $D$-cocycles $\omega_{D}:G^D\to U(1)$ that maps $D$ independent elements of $G$ to a $U(1)$ phase, and satisfies the cocycle condition that its differential map is trivial:
\begin{equation*}
    1=(d_D\omega_D)(g_1,g_2,\dots,g_{D+1})\equiv\omega_D^{s(g_1)}(g_2,\dots,g_{D+1})\prod_{i=1}^D \omega_D^{(-1)^i}(g_1,\dots,g_{i-1},g_ig_{i+1},g_{i+2},\dots)\omega_D^{(-1)^{D+1}}(g_1,\dots,g_D)
\end{equation*}
where $s(g)=-1$ when $g$ is time-reversal or other anti-unitary symmetry, and otherwise $s(g)=1$. Two cocycles $\omega_D$ and $\omega_D'$ are equivalent if they differ only by a co-boundary:
\begin{equation}
    \frac{\omega_D(g_1,\dots,g_D)}{\omega_D'(g_1,\dots,g_D)}=d_{D-1}w_{D-1}(g_1,\dots,g_D),
\end{equation}
where $w_{D-1}:G^{D-1}\to U(1)$ is a $(D-1)$-cocycle. For later convenience, we define the cochain $\nu_{D}:G^{D+1}\to U(1)$, so that
 \begin{equation}
     \nu_{D}(1,g_1,g_2,\dots,g_D)=\omega_D(g_1,g_1^{-1}g_2,g_2^{-1}g_3,\dots,g_{D-1}^{-1}g_D),
 \end{equation}
and \beqn\nu(1,g_1,\dots,g_D)=\nu^{s(g)}(g,gg_1,\dots,gg_D)\eeqn for any $g\in G$. The cochain satisfies a simpler cocycle condition
\begin{equation}
    1=d_D\nu_D(g_0,g_1,\dots,g_{D+1})=\prod_{i=0}^{D+1}\nu_D^{(-1)^i}(g_0,\dots,g_{i-1},g_{i+1},\dots,g_{D+1}).
\end{equation}

The wave function of the corresponding SPT state can be directly constructed from the $D$-cocycle $\omega_D$. The $D-1$ dimensional space is first triangularized into simplices, and a branching rule is assigned so that one can assign an order of the $D$ vertices in the simplex $S$ as $S_1,S_2,\dots,S_{D}$. Each simplex is also assigned with an orientation $\epsilon_S=\pm 1$. The local Hilbert space on each lattice site $i$ is spanned by a basis form by all the group elements, $\{\ket{g_i}|g\in G\}$. The symmetry acts locally as $U_{g}\ket{g_i}=\ket{g g_i}$ for any $g\in G$. The SPT wave function can be constructed by a linear superposition of all $g$ configurations:
\begin{equation*}
    \ket{\psi_\text{SPT}}\sim \sum_{\{g_i\}}\prod_s \nu^{\epsilon_S}_{D}(1,g_{S_1},g_{S_2},\dots,g_{S_{D}})\bigotimes_i\ket{g_i}.
\end{equation*}
The wave function is symmetric under $G$ actions, since any symmetry action $U_g\ket{\psi_\text{SPT}}$ on the basis vectors $\ket{g_i}$ in $\ket{\psi_\text{SPT}}$ can be absorbed into the factors of the simplices, and each factor now becomes
\begin{equation}
\begin{aligned}
    \nu_D^{s(g)\epsilon_S}(1,g^{-1}g_{S_1},\dots,g^{-1}g_{S_D})&=\nu_D^{\epsilon_S}(g,g_{S_1},\dots,g_{S_D})\\
    &=\nu_D^{\epsilon_S}(1,g_{S_1},\dots,g_{S_D})\prod_{i=1}^D\nu_D^{\epsilon_S(-1)^{i}}(1,g,g_{S_1},\dots,g_{S_{i-1}},g_{S_{i+1}},\dots,g_{S_D}),
\end{aligned}
\end{equation}
where we have used the cocycle condition in the last equality. On an infinite system, the extra factors in the product above will be cancelled by nearby simplices that share the $D-2$ dimensional simplex with $S$, hence the entire product is invariant. However, on an open lattice, the symmetry action will leave extra factors on the lattice boundary, so that the wave function will no longer be symmetric:
\begin{equation}
    U_g\ket{\psi_\text{SPT}}\sim \sum_{\{g_i\}} \prod_{S\in\partial}\nu_D^{\epsilon_\mathfrak{S}}(1,g,g_{\mathfrak{S}_1},\dots,g_{\mathfrak{S}_{D-1}})\prod_s \nu^{\epsilon_s}_{D}(1,g_{s_1},g_{s_2},\dots,g_{s_{D}})\bigotimes_i\ket{gg_i}.
\end{equation}
Here $S\in\partial$ means that one $D-1$ dimensional face $\mathfrak{S}$ of the simplex $S$ is on the lattice boundary. Hence, the symmetry action on the spins on the boundary can be effectively written as
\begin{equation}
    U_g\bigotimes_{i\in\partial}\ket{g_i}=\prod_{\mathfrak{S}}\nu_D^{\epsilon_\mathfrak{S}}(1,g,g_{\mathfrak{S}_1},\dots,g_{\mathfrak{S}_{D-1}})\bigotimes_{i\in\partial}\ket{gg_i}=\prod_{\mathfrak{S}}\omega_D^{\epsilon_\mathfrak{S}}(g,g^{-1}g_{\mathfrak{S}_1},g_{\mathfrak{S}_1}^{-1}g_{\mathfrak{S}_2},\dots,g_{\mathfrak{S}_{D-2}}^{-1}g_{\mathfrak{S}_{D-1}})\bigotimes_{i\in\partial}\ket{gg_i}
    \label{symact}
\end{equation}

The anomalous symmetry actions on the boundary of various SPT states in the main text can be directly obtained from the result above. For example, for the Levin-Gu state protected by $\Z_2=\{I,g\}$, the only non-trivial map of the 3-cocycle is $\omega_3(g,g,g)=-1$, while any other combination of three group elements is mapped to $1$. If one defines $\ket{I}=\ket{\sigma^z=1}$ and $\ket{g}=\ket{\sigma^z=-1}$, the symmetry action on the 1+1$D$ boundary can be obtained from \eqnref{symact} as (see can simply set $\epsilon_\mathcal{S}=1$ here)
\begin{equation}
    U_g\bigotimes_{i=1}^{N_\partial}\ket{\sigma^z_i=n_i}=\prod_{i=1}^{N_\partial}\omega_3\left(g,gg^{\frac{1-n_i}{2}},g^{\frac{1-n_i}{2}}g^{\frac{1-n_{i+1}}{2}}\right)\bigotimes_{i=1}^{N_\partial}\ket{\sigma^z_i=-n_i}=\prod_{i=1}^{N_\partial}C(n_i,n_in_{i+1})\bigotimes_{i=1}^{N_\partial}\ket{\sigma^z_i=-n_i}.
\end{equation}
Hence $U_g=\prod_{i=1}^{N_\partial}C(\sigma^z_i,\sigma^z_i\sigma^z_{i+1})\sigma^x_i$, which agrees with the decorated defect result in \secref{sec:anomaly}. 

\begin{figure}
    \centering
    \includegraphics[width=4.5cm]{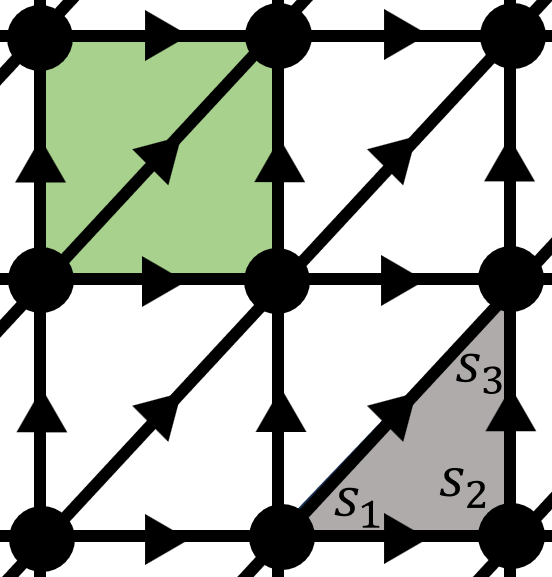}
    \caption{The triangular lattice in writing down the anomalous symmetry action of $U_K$. The black arrows indicate the branching rule of each triangular plaquette. Two nearby triangles (marked in green) can be combined so that the controlled unitaries can be written on a square lattice in \eqnref{UAUK2}.}
    \label{fig:tritosq}
\end{figure}

The symmetry action of the $G=\Z_2^2$-anomalies $\Omega_{1,2}$ in 2+1$D$ can also be obtained in this way. Denoting the non-trivial element of $A$ and $K$ by $a$ and $k$, the anomaly $\Omega_1$ is characterized by the following 4-cocycle (for $\Omega_2$ the role of $a$ and $k$ is simply reversed):
\begin{equation}
    \omega_4(a^ik,a^jk,a^lk,ak^m)=-1
\end{equation}
for any $i,j,l,m=0$ or $1$, while all the other cocycles are equal to $1$. To write down the anomalous symmetry actions, we define two sets of Pauli spins $\sigma$ and $\tau$, so that $\ket{a^ik^l}=\ket{\sigma^z=1-2i}\otimes\ket{\tau^z=1-2l}$. Since the values of the cocycles are $\pm 1$, the orientation does not matter, and the symmetry actions on a triangular lattice can be written as
\begin{equation}
\begin{aligned}
    U_A\bigotimes_i\ket{\sigma^z_i=m_i}\otimes\ket{\tau^z_i=n_i}=&\ket{\sigma^z_i=-m_i}\otimes\ket{\tau^z_i=n_i},\\
    U_K\bigotimes_i\ket{\sigma^z_i=m_i}\otimes\ket{\tau^z_i=n_i}=&\prod_{\triangle_s}\omega_4(k,kk^{\frac{1-n_{s_1}}{2}},k^{\frac{1-n_{s_1}}{2}}k^{\frac{1-n_{s_2}}{2}},a^{\frac{1-m_{s_2}}{2}}a^{\frac{1-m_{s_3}}{2}})\bigotimes_i\ket{\sigma^z_i=m_i}\otimes\ket{\tau^z_i=-n_i}\\
    =&\prod_{\triangle_s}C(-n_{s_1},n_{s_1}n_{s_2},m_{s_2}m_{s_3})\bigotimes_i\ket{\sigma^z_i=m_i}\otimes\ket{\tau^z_i=-n_i}\\
    =&\left[\prod_{\triangle_s}C(\tau^z_{s_1},\tau^z_{s_1}\tau^z_{s_2},\sigma^z_{s_2}\sigma^z_{s_3})\prod_i\tau^x_i\right]\bigotimes_i\ket{\sigma^z_i=m_i}\otimes\ket{\tau^z_i=n_i}.
\end{aligned}
\label{UAUKcohomology}
\end{equation}
To recover the convenient forms of anomalous symmetry actions in \eqnref{UAUK2}, we combine controlled unitaries of two nearby triangles with opposite orientations (see \figref{fig:tritosq}). Rearranging the controlled unitaries in each plaquette, it is straightforward to verify that the non-on-site factor of $U_K$ in \eqnref{UAUKcohomology} is equal to the one in \eqnref{UAUK2}.

In a similar fashion, the 1+1$D$ $\Z_2^2$ anomaly discussed in \secref{sec:stwmodel} can also be obtained from the non-trivial 3-cocycle $\omega_3(a^ik,a^jk,ak^l)=-1$ for any $i,j,l=0,1$ using two sets of Pauli spins $\sigma$ and $\tau$.

\section{Construction of the symmetric infinite temperature mixed state}
\label{app:derivation}

In this appendix, we carry out the calculation in the first line of \eqnref{inftemp} and the explicit form of the random Hamiltonian in \eqnref{HJh}.

\subsection{The explicit form of the random Hamiltonian in \eqnref{HJh} }
We now compute the random unitary conjugation in \eqnref{HJh} explicitly. It is easy to see that the $J_\sigma$ terms are invariant under the action of $W(\{h_i,l_i\})$. Meanwhile, the $J_\tau$ terms transform under conjugation of $V_i$ as
\begin{equation}
    V_i\tau^y_{i+\frac{1}{2}}\left(1+\sigma^z_i\sigma^z_{i+1}\right)V_i^{-1}=\tau^y_{i+\frac{1}{2}}\left(-\sigma^z_i+\sigma^z_{i+1}\right)=-V_{i+1}\tau^y_{i+\frac{1}{2}}\left(1+\sigma^z_i\sigma^z_{i+1}\right)V_{i+1}^{-1}.
\end{equation}
Other conjugations by $V$ can be similarly computed. Using the fact that the unitary $V_i$ commutes with $V_j$ for any sites $i$ and $j$, and conjugating by $\tau^z$ only flips the sign of the $J_\tau$ term, we collect all the conjugation results and arrive at the explicit form of the disordered Hamiltonian:
\begin{equation}
    H(\{h_i,l_i\})=-\sum_i\left[J_\tau l_i\frac{1+h_ih_{i+1}}{2}\tau^y_{i+\frac{1}{2}}\left(1+\sigma^z_i\sigma^z_{i+1}\right)+J_\tau l_i\frac{h_{i+1}-h_{i}}{2}\tau^y_{i+\frac{1}{2}}\left(\sigma^z_i-\sigma^z_{i+1}\right)+J_\sigma\sigma^x_{i}\left(1+\tau^z_{i-\frac{1}{2}}\tau^z_{i+\frac{1}{2}}\right)\right].
\end{equation}
We see that $H(\{h_i,l_i\})$ is essentially a random-bond Ising model whose bond-interaction  is charged under $A$. Therefore, the bond-randomness breaks $A$ down to an average while preserving $K$ exactly.

\subsection{Derivation of \eqnref{inftemp}}
We now derive the density matrix that describes the ensemble of disordered ground states of $H(\{h_i,l_i\})$. Using the definition in \eqnref{Whl}, the pure ferromagnetic state $\ket{\psi_0}$ is transformed into

\begin{equation}
\begin{aligned}    W(\{h_i,l_i\})\ket{\psi_0}=&\prod_i \left[\sigma^x_iC\left(\sigma^z_i,\tau^z_{i-\frac{1}{2}}\tau^z_{i+\frac{1}{2}}\right)\right]^{\frac{1-h_i}{2}}\left(\tau^z_{i+\frac{1}{2}}\right)^{\frac{1-l_i}{2}}\frac{1}{\sqrt{2}}\left(\ket{\uparrow\uparrow\dots}_\sigma+\ket{\downarrow\downarrow\dots}_\sigma\right)\otimes\ket{+_y+_y\dots}_\tau\\
=&\frac{\bigotimes_i\ket{\sigma^z_i=h_i}\ket{\tau^y_{i+\frac{1}{2}}=l_i}+\bigotimes_i\ket{\sigma^z_i=-h_i}\ket{\tau^y_{i+\frac{1}{2}}=l_ih_ih_{i+1}}}{\sqrt{2}}\\
=&\frac{\mathbbm{1}+U_K}{\sqrt{2}}\bigotimes_i\ket{\sigma^z_i=h_i}\ket{\tau^y_i=l_i}.
\end{aligned}
\end{equation}
In the last equality we have used the definition of $U_K$ from \eqnref{UAUK}. Collecting all possible values of $h_i,l_i=\pm1$, we have
\begin{equation}
    \rho_{\Z_2^2}=\frac{\mathbbm{1}+U_K}{\sqrt{2}}\frac{\mathbbm{1}}{2^{2N}}\frac{\mathbbm{1}+U_K}{\sqrt{2}}=\frac{\mathbbm{1}+U_K}{2^{2N}}.
\end{equation}

There is an easier derivation by noticing that the clean ferromagnetic state is stabilized these following $2N$ unitaries: $\sigma^z_i\sigma^z_{i+1}$, $\tau^z_{i+\frac{1}{2}}$, and $U_K$. Unitaries in $W(\{h_i,l_i\})$ randomly flip the eigenvalues of every local stabilizer between $\pm1$, but keep $U_K=1$. Hence, the eventual density matrix is proportional to the projection operator $\frac{\mathbbm{1}+U_K}{2}$.

\section{The derivation of the unitary operators $\mathcal{V}$ and $\mathcal{W}$ via group extension and gauging on the lattice}
\label{app:VW}
As outlined in \secref{sec:SET} and \secref{sec:z22model}, the TO that carries the exact anomaly $\Omega_1$ of $G=A\times K=\Z_2^2$ can be constructed by first extending the global symmetry to $G'=D_4$ via the short exact sequence $1\to N=\Z_2\to D_4\to G=\Z_2^2\to 1$, after which the $G$-anomaly will be trivialized. This means that the $D_4$ symmetry should act in a completely on-site way, which can be represented using three sets of Pauli operators on a 2$D$ square lattice in the following way:
\begin{equation}
    \tilde U_A=\prod_i\sigma^x_i,\ \tilde U_K=\prod_i C(\mu^x_{i},\sigma^z_i)\tau^x_i,\ \tilde U_N=\prod_i\mu^x_i.
    \label{D4action}
\end{equation}
The group formed by the three symmetry actions is indeed $D_4$, since $(\tilde U_A\tilde U_K)^2=\tilde U_N$. Our goal is to gauge the normal subgroup $N$ and recover the anomalous symmetry actions of $G$ in \eqnref{UAUK2}.

As a warm-up, we first review the usual way that the $\Z_2$ global symmetry is gauged on the lattice. This will result in an anomaly-free $\Z_2$ gauge theory enriched by $G=A\times K=\Z_2^2$, where the $e$ anyon (the gauge charge of $N$) carries a projective representation of $A$ and $K$. To this end, we introduce the $\Z_2$ gauge field of $N$ on each link $l$ of the lattice, whose Pauli operators are denoted by $a_l^{x,z}$ (see \figref{fig:z22lattice}). The local gauge constraint is implemented via the Gauss law, whose form on the lattice reads
\begin{equation}
    \mu^x_i\prod_{i\in\partial l}a^x_l\equiv \mu^x_i\mathcal{A}_i=1
    \label{gauss}
\end{equation}
for every lattice site $i$. The local $N$ charge excitation is gauge-invariant only if being attached by a Wilson line: $\mu^z_i\left(\prod_{l\in\gamma_{ij}}a^z_l\right) \mu^z_j$, where $\gamma_{ij}$ is a line that connects sites $i$ and $j$. In the deconfined phase, each plaquette has vanishing $\Z_2$ gauge flux, so that $\mathcal{B}_p\equiv\prod_{l\in p}a^z_l=1$ for every $p$. This means that we can effectively replace $\mu^z_{l_1}\mu^z_{l_2}$ by $a^z_l$, where $l_{1,2}$ are the two neighboring lattice sites connected by $l$.

We now gauge $N$ in the extended symmetry action in \eqnref{D4action} following the recipe above. Replacing $\mu^x_i$ in $U_K$ using the Gauss law, we arrive at the following $\Z_2^2$ global symmetry action
\begin{equation}
    \tilde U_A\to\prod_i\sigma^x_i,\  \tilde U_K\to\prod_iC\left(\prod_{i\in\partial l}a^x_l,\sigma^z_i\right)\tau^x_i=\mathcal{W}\prod_i\tau^x_i\mathcal{W},
\end{equation}
where the unitary operator $\mathcal{W}$ is given by \eqnref{calW}. Therefore, in the $\mathcal{W}$-transformed frame, the global symmetry is completely on site: the global symmetry action of $K$ is simply $\prod_i\tau^x_i$, and $U_A=\prod_i\sigma^x_i$ is symmetric under conjugation of $\mathcal{W}$. The $\Z_2$ gauge theory of $N$ in the $\mathcal{W}$-transformed frame is therefore decoupled from the matter fields $\sigma$ and $\tau$.

To obtain the anomalous symmetry action $U_K$ in \eqnref{UAUK2}, the gauging of $N$ needs to be done in a ``twisted" way\cite{tachikawa2020gauging}, so that the $N$ charge is tied to the local configuration of the $\sigma$ and $\tau$ spins. More explicitly, the Gauss law of the twisted gauging should be enforced as
\begin{equation}
    \mu^x_i \prod_{i\in\partial l}a^x_{l}=C(-\tau^z_{p_1}\tau^z_{p_3},\tau^z_{p_2}\tau^z_{p_4}),
    \label{twistedgauss}
\end{equation}
where the plaquette $p$ is located at the bottom-left of the site $i$. Substituting the relation above into $\tilde U_K$ in \eqnref{D4action}, we indeed recover the anomalous symmetry action of $K$ in \eqnref{UAUK2}:
\begin{equation}
    \tilde U_K\to \prod_i C\left(\prod_{i\in\partial l}a^x_{l},\sigma^z_i\right)\prod_p C(-\tau^z_{p_1}\tau^z_{p_3},\tau^z_{p_2}\tau^z_{p_4},\sigma^z_{p_3})\prod_i\tau^x_i=\mathcal{W}U_K\mathcal{W}^{-1}.
\end{equation}

Finally, we note that the twisted Gauss law in \eqnref{twistedgauss} can actually be ``untwisted" using the unitary operator $\mathcal{V}$ defined in \eqnref{calV}:
\begin{equation}
\begin{aligned}
    \mathcal{V}\left(\mu^x_i \prod_{i\in\partial l}a^x_{l}\right)\mathcal{V}^{-1}&=\mu^x_i\prod_{i\in\partial l}a^x_{l}\times C(\tau^z_{p_1},\tau^z_{p_1}\tau^z_{p_2})C(\tau^z_{p_1},\tau^z_{p_1}\tau^z_{p_4})C(\tau^z_{p_2},\tau^z_{p_2}\tau^z_{p_3})C(\tau^z_{p_4},\tau^z_{p_4}\tau^z_{p_1})\\
    &=\mu^x_i \prod_{i\in\partial l}a^x_{l}\times C(-\tau^z_{p_1}\tau^z_{p_3},\tau^z_{p_2}\tau^z_{p_4})=1.
\end{aligned}
\end{equation}
Therefore, in the $\mathcal{V}\mathcal{W}$-transformed frame, the gauge sector is completely decoupled from the matter fields $\sigma$ and $\tau$, and the wave function of the entire system should be a product state of the form in \eqnref{psischematic}.
\twocolumngrid
\bibliographystyle{unsrt}
\bibliography{paper}

\end{document}